\documentclass[final,5p,times,twocolumn]{elsarticle}

\usepackage{amssymb}
\usepackage{amsmath}
\usepackage{amsthm}
\usepackage{bm}
\usepackage{multirow}
\usepackage{makecell}
\usepackage{adjustbox}
\usepackage{booktabs}
\usepackage{textcomp}
\usepackage{hyperref}
\hypersetup{
    colorlinks=true,
    linkcolor=blue,
    filecolor=gray,      
    urlcolor=blue,
    citecolor=blue,
}

\newcounter{bla}

\journal{Computer Physics Communications}

\begin{document}

\begin{frontmatter}
\title{DeepH-pack: A general-purpose neural network package for deep-learning electronic structure calculations}

\author[thupys,thulab]{Yang Li\fnref{eq1}}
\author[thupys,stfmatsci]{Yanzhen Wang\fnref{eq1}}
\author[thupys,thulab]{Boheng Zhao\fnref{eq1}}
\author[thupys,caphys,caphyslib]{Xiaoxun Gong}
\author[thupys,thulab,thuphysadv]{Yuxiang Wang}
\author[thupys,thulab]{Zechen Tang}
\author[thupys,thulab]{Zixu Wang}
\author[thupys,thulab]{Zilong Yuan}
\author[thupys,thulab]{Jialin Li}
\author[thupys,thulab]{Minghui Sun}
\author[thupys,thulab]{Zezhou Chen}
\author[thupys,thulab]{Honggeng Tao}
\author[thupys,thulab]{Baochun Wu}
\author[thupys,thulab]{Yuhang Yu}
\author[thupys,thulab,thuphysadv]{He Li}
\author[stfmatsci]{Felipe H. da Jornada\corref{corref2}}
\author[thupys,thulab,thuphysadv,thuqi]{Wenhui Duan\corref{corref1}}
\author[thupys,thulab,thuqi]{Yong Xu\corref{corref1}}

\affiliation[thupys]{
    organization={Department of Physics},
    addressline={Tsinghua University}, 
    city={Beijing},
    postcode={100084}, 
    state={Beijing},
    country={China}
}
\affiliation[thulab]{
    organization={State Key Laboratory of Low Dimensional Quantum Physics},
    addressline={Tsinghua University}, 
    city={Beijing},
    postcode={100084},
    state={Beijing},
    country={China}
}
\affiliation[thuphysadv]{
    organization={Institute for Advanced Study},
    addressline={Tsinghua University}, 
    city={Beijing},
    postcode={100084}, 
    state={Beijing},
    country={China}
}
\affiliation[thuqi]{
    organization={Frontier Science Center for Quantum Information},
    addressline={Tsinghua University}, 
    city={Beijing},
    postcode={100084}, 
    state={Beijing},
    country={China}
}
\affiliation[stfmatsci]{
    organization={Department of Materials Science and Engineering},
    addressline={Stanford University}, 
    city={Stanford},
    postcode={94305}, 
    state={CA},
    country={USA}
}
\affiliation[caphys]{
    organization={Department of Physics},
    addressline={University of California}, 
    city={Berkeley},
    postcode={94720}, 
    state={CA},
    country={USA}
}
\affiliation[caphyslib]{
    organization={Materials Sciences Division},
    addressline={Lawrence Berkeley National Laboratory}, 
    city={Berkeley},
    postcode={94720}, 
    state={CA},
    country={USA}
}

\cortext[corref1] {Corresponding authors at: Department of Physics, Tsinghua University, Beijing, 100084, Beijing, China. \textit{E-mail:} yongxu@mail.tsinghua.edu.cn, duanw@tsinghua.edu.cn}

\cortext[corref2] {Corresponding author at: Department of Materials Science and Engineering, Stanford
University, Stanford, 94305, CA, USA.\\\textit{E-mail:} jornada@stanford.edu}

\fntext[eq1]{These authors contributed equally to this work.}

\begin{abstract}
In computational physics and materials science, first-principles methods, particularly density functional theory, have become central tools for electronic structure prediction and materials design. Recently, rapid advances in artificial intelligence (AI) have begun to reshape the research landscape, giving rise to the emerging field of deep-learning electronic structure calculations. Despite numerous pioneering studies, the field remains in its early stages; existing software implementations are often fragmented, lacking unified frameworks and standardized interfaces required for broad community adoption. Here we present DeepH-pack, a comprehensive and unified software package that integrates first-principles calculations with deep learning. By incorporating fundamental physical principles into neural-network design, such as the nearsightedness principle and the equivariance principle, DeepH-pack achieves robust cross-scale and cross-material generalizability. This allows models trained on small-scale structures to generalize to large-scale and previously unseen materials. The toolkit preserves first-principles accuracy while accelerating electronic structure calculations by several orders of magnitude, establishing an efficient and intelligent computational paradigm for large-scale materials simulation, high-throughput materials database construction, and AI-driven materials discovery.
\\


\noindent \textbf{PROGRAM SUMMARY}

\begin{small}
\noindent
{\em Program Title:} DeepH-pack\\
{\em Licensing provisions:} GPLv3\\
{\em Programming language: Python}\\
{\em Nature of problem:} Developing efficient and intelligent first-principles methods using deep learning.\\
{\em Solution method:} DeepH-pack employs neural networks implemented in the deep learning framework JAX to model the input–output relationships of first-principles calculations, integrating fundamental physical constraints into the network design to improve generalizability and computational efficiency of the deep learning approach.\\
{\em Additional comments including restrictions and unusual features:} The code defines a unified data specification for storing key physical quantities (such as Hamiltonian, overlap matrices, charge density, and density matrices) in a standardized format. This specification is interfaced with mainstream density functional theory software packages, enabling straightforward and efficient data exchange between DeepH-pack and these third-party codes.\\
\end{small}
\end{abstract}
\end{frontmatter}

\section{Introduction}

Density functional theory (DFT) has established itself as a standard method for first-principles calculations, playing a pivotal role in understanding complex quantum materials and designing novel devices~\cite{hohenberg1964inhomogeneous,kohn1965self}. Despite its success, the computational cost of conventional DFT scales cubically with the number of electrons, becoming prohibitive for large systems. This computational complexity fundamentally restricts the applicability of DFT in demanding tasks, such as large-scale molecular dynamics, electronic-structure simulations, and high-throughput screening across massive materials databases~\cite{martin2020electronic}. Consequently, the trade-off between accuracy and efficiency poses a persistent bottleneck for realistic device modeling and materials discovery. In response to this challenge, artificial intelligence (AI), particularly deep learning, has emerged as a transformative paradigm. By training on high-quality data generated by first-principles calculations, deep neural networks can capture the underlying quantum-mechanical mappings from atomic structures to physical properties, offering the potential to achieve prediction accuracy comparable to reference first-principles methods with significantly reduced computational cost.

The remarkable success of deep learning across diverse scientific domains has accelerated its adoption in \emph{ab initio} materials simulation~\cite{merchant2023scaling,zeni2025generative,szymanski2023autonomous,unke2021machine,schutt2019unifying,li2022deep}. While earlier efforts focused primarily on developing machine learning interatomic potentials (MLIPs) to predict energy and forces~\cite{xie2018crystal,schutt2018schnet,zhang2018deep,unke2021se}, the frontier has recently expanded to direct prediction of electronic structure quantities~\cite{tang2025deep,li2025critical}. There is a growing focus on using neural networks to reproduce high-dimensional, multi-faceted properties with quantum-mechanical origins, including charge densities~\cite{grisafi2018transferable,koker2024higher}, density matrices~\cite{shao2023machine,tang2024improving}, wave functions~\cite{pfau2020ab,hermann2020deep}, and band structures~\cite{nunez2019exploring}. These developments allow machine-learning-assisted simulations to go beyond thermodynamic and mechanical properties, enabling the investigation of a wider range of electronic, magnetic, and optical phenomena.

Among these electronic quantities, the DFT Hamiltonian holds a central position. As the fundamental operator governing quantum material systems, the DFT Hamiltonian encodes essential information about the electronic structure properties. Once the Hamiltonian is determined, in principle all physical observables at the DFT level, such as band structures, response functions, and transport properties, can be derived through standard post-processing calculations~\cite{martin2020electronic,holst2011electronic}. Motivated by this capability, the deep-learning DFT Hamiltonian (DeepH) method pioneered the direct prediction of DFT Hamiltonians via deep learning~\cite{li2022deep}. As the first approach to generalize from small-scale material systems to large ones and achieve strict quantum-mechanical fidelity in this domain, DeepH exemplifies the \emph{Physics for AI} paradigm: It leverages fundamental physical principles, including the nearsightedness principle of electronic matter and the equivariance principle, to guide the design of neural network architecture and the selection of target quantities for deep learning. This physics-informed design ensures high data efficiency and extraordinary generalizability of the DeepH approach, as demonstrated in the original work, which provided the first open-source code addressing this problem~\cite{li2022deep}. At present, several related algorithms have been developed, including the series of DeepH algorithms~\cite{gong2023general,li2023deep}, QHNet~\cite{yu2023efficient},  HamGNN~\cite{su2023efficient,zhong2023transferable}, Hot-Ham~\cite{liang2025hot}, among others~\cite{yin2024towards}. By successfully embedding physical priors into deep-learning models, DeepH establishes a virtuous cycle, employing AI as a reliable tool to accelerate and expand the frontiers of physical research, leading to various applications in electronic structure analysis~\cite{liu2024breaking,yang2024evolution,huang2025atlas,zhu2025predicting,ke2025combining}.

Building on this methodological foundation, DeepH has evolved into a general-purpose algorithmic framework with high versatility and scalability. It is capable of modeling systems across various length scales, ranging from small molecules to complex solids containing tens of thousands of atoms, while handling intricate physical effects such as spin-orbit coupling (SOC) and magnetic ordering. Furthermore, the framework's applicability extends well beyond the sole prediction of the DFT Hamiltonian. When paired with corresponding data flows, the DeepH method can be generalized to predict a wide array of physical quantities that transform as scalars or tensors, including energies and atomic forces~\cite{yuan2024equivariant}, density matrices~\cite{tang2024improving}, real-space Kohn-Sham potentials~\cite{yuan2024deep}, charge densities, as well as physical quantities within the generalized Kohn-Sham framework~\cite{tang2023efficient} and density functional perturbation theory~\cite{li2024deep}. Operationally, it allows for seamless integration with existing DFT software ecosystems, where the predicted quantities can serve as high-quality initial guesses to accelerate self-consistent field (SCF) convergence or be used directly for post-processing tasks. This establishes a symbiotic workflow that combines the efficiency of deep learning with the rigorous formalism of first-principles calculations.

While significant progress has been made at the algorithmic level~\cite{li2022deep,gong2023general,li2023deep,yuan2024equivariant,tang2023efficient,li2024deep,wang2024deeph,li2024neural}, realizing AI-driven materials discovery still requires substantial foundational engineering efforts in software and implementation. Bridging the gap between methodological advances and practical applications demands more than theoretical and algorithmic sophistication: it critically depends on the availability of high-performance, scalable, and user-friendly software implementations. Although promising approaches for predicting key electronic properties continue to emerge, the development of a unified deep learning framework, characterized by versatility, scalability, and ease of use across diverse material systems and length scales, would greatly benefit the research community. Such a framework has the potential to significantly accelerate the adoption of these advanced methods in practical \emph{ab initio} simulations.

\begin{figure}
    \centering
    \includegraphics[width=\linewidth]{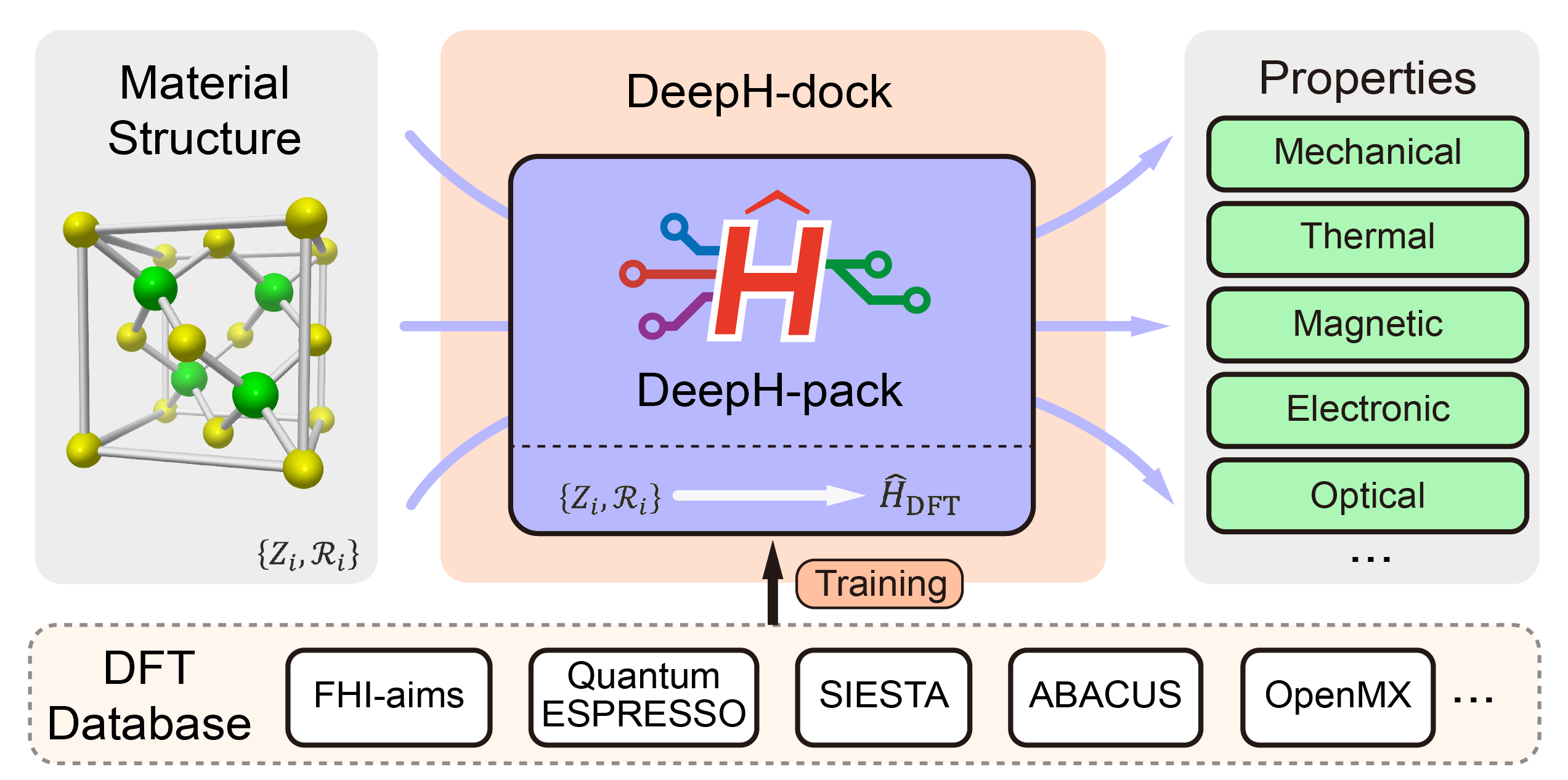}
    \caption{Schematic of the DeepH computational workflow. Starting from the atomic structure, DeepH-pack predicts the material's DFT Hamiltonian, enabling the prediction of various material properties with first-principles accuracy. DeepH-dock serves as the standardized data interface, converting data from various computational software packages (e.g., FHI-aims~\cite{blum2009ab}, Quantum ESPRESSO~\cite{giannozzi2009quantum}, SIESTA~\cite{soler2002siesta}, ABACUS~\cite{zhou2025abacus}, OpenMX~\cite{ozaki2003variationally}, \emph{etc}.) for model training. DeepH-pack constitutes the core algorithmic component of the software, handling both model training and inference using existing models.}
    \label{fig1}
\end{figure}

In this article, we present DeepH-pack, a software package implementing neural network models that are designed to predict DFT Hamiltonians for atomistic systems. DeepH-pack provides an implementation of the DeepH method, which uses advanced graph neural network (GNN) architectures and symmetry-aware methods to predict DFT Hamiltonian matrices from atomic configurations. The time complexity of the method grows linearly with the number of atoms, greatly improving upon the cubic complexity that is typical of traditional DFT methods. Furthermore, even for small systems with only tens of atoms, DeepH-pack predicts Hamiltonians two to three orders of magnitude faster than standard DFT calculations. Despite its substantial speedup, DeepH-pack maintains very high accuracy, achieving sub-meV agreement with conventional DFT methods in electronic band structures. To ensure broad compatibility, DeepH-pack provides automated interfaces for major DFT software packages, including FHI-aims~\cite{blum2009ab,abbott2025roadmap}, Quantum ESPRESSO~\cite{giannozzi2009quantum}, SIESTA~\cite{soler2002siesta}, ABACUS~\cite{zhou2025abacus}, and OpenMX~\cite{ozaki2003variationally}. This compatibility allows users to integrate DeepH-pack easily into existing computational workflows. By supporting both crystalline solids and molecular systems, DeepH-pack serves as a versatile and powerful tool for the computational materials science community. The computational workflow of DeepH-pack is schematically displayed in Fig. \ref{fig1}.

\section{Methods}
\subsection{Theoretical Background}
The main capability of DeepH-pack is to predict the DFT Hamiltonian $\hat{H}_{\mathrm{DFT}}$, a core quantity in the Kohn-Sham formalism~\cite{hohenberg1964inhomogeneous}. In this theory, $\hat{H}_{\mathrm{DFT}}$ is an effective one-electron operator that serves as a compact gateway to many properties of interest at the DFT level. From its eigenvalues $\{\mathcal{E}_n\}$ and eigenstates $\{|\psi_n\rangle \}$, one can compute the electron density, total energy, Hellmann-Feynman forces, and, for periodic crystals, the band structure and the density of states (DOS). Practically, $\hat{H}_{\mathrm{DFT}}$ is a cornerstone for deriving many properties of interest, including charge and spin densities, forces and stresses, and a variety of transport and response functions~\cite{scheidemantel2003transport,gajdovs2006linear,rohlfing2000electron}. Therefore, DeepH-pack focuses on predicting the DFT Hamiltonian as a fundamental step toward enabling accurate and efficient electronic structure modeling (Fig. \ref{fig2}(a)).

In conventional DFT calculations, to obtain the Hamiltonian, one needs to solve the Kohn-Sham equation
\begin{equation}
\label{dfteig}
    \hat{H}_{\text{DFT}}\left|\psi_n \right\rangle=\mathcal{E}_n \left|\psi_n \right\rangle
\end{equation} 
self-consistently, which is time-consuming due to its cubic complexity in the number of atoms and the cumbersome self-consistent iterations. However, a more straightforward route exists with the help of AI. According to the Hohenberg-Kohn theorem, $\hat{H}_{\text{DFT}}$ is uniquely determined by the material structure, i.e., the atomic species $\left\{ Z_i \right\}$ and the atomic positions $\left\{ \mathcal{R}_i \right\}$~\cite{hohenberg1964inhomogeneous}. Therefore, DeepH-pack aims to model a direct mapping $\left\{ Z_i, \mathcal{R}_i \right\} \mapsto \hat{H}_{\text{DFT}}\left( \left\{ Z_i, \mathcal{R}_i \right\} \right)$ using deep learning. To make full use of the locality principle, DeepH-pack follows the original work of Li \emph{et al.}~\cite{li2022deep} and predicts the matrix elements. Given atoms $i$ and $j$ located at positions $\mathcal{R}_i$ and $\mathcal{R}_j$ within their respective unit cells, which are displaced by lattice vectors $\bm{R}_i$ and $\bm{R}_j$, the Hamiltonian matrix elements are defined as:
\begin{equation}
    H_{i\alpha, j\beta}(\bm{R}_j-\bm{R}_i) = \int d^3 \bm{r} d^3 \bm{r^{\prime}} \phi^{*}_{i\alpha}(\bm{r}-\mathcal{R}_i) \hat{H}_{\text{DFT}}\phi_{j\beta}(\bm{r}^{\prime}-\mathcal{R}_j)
\end{equation}
under a basis of localized atomic-orbital functions $\left\{\phi_{i\alpha}\right\}$. Here, $\alpha,\beta$ label orbitals on those atoms. When SOC is present, $\alpha$ and $\beta$ have an additional spin degree of freedom. In the remaining sections of the paper, we use $H_{ij}$ to denote the sub-blocks of Hamiltonian matrix elements associated with the atom-pair $i$ and $j$.

After obtaining the DFT Hamiltonian, one can solve the non-self-consistent Kohn-Sham equation (\ref{dfteig}) to obtain its eigenspectrum for further study. For periodic systems, Eq. (\ref{dfteig}) takes the form of a generalized eigenvalue problem~\cite{wang2019first}:
\begin{equation}
\label{aoeig}
    \sum_{j\beta \in \mathrm{uc}} H_{i\alpha, j\beta}(\bm{k}) c_{n, j\beta}(\bm{k}) = \mathcal{E}_n(\bm{k}) \sum_{j\beta \in \mathrm{uc}} S_{i\alpha, j\beta}(\bm{k}) c_{n, j\beta}(\bm{k}),\ i\alpha \in \mathrm{uc},
\end{equation}
$\mathcal{E}_n(\bm{k})$ are called the electronic band structure, and $c_{n, j\beta}(\bm{k})$ are the wave function coefficients in the given basis. Here, ``uc'' denotes the unit cell, and the matrix elements in reciprocal space are constructed from the real-space Hamiltonian and overlap matrices via the Fourier transform:
\begin{equation}
O_{i\alpha, j\beta}(\bm{k}) = \sum_{\bm{R}} e^{i \bm{k} \cdot \bm{R}} O_{i\alpha, j\beta}(\bm{R}), \quad O \in \{H, S\},
\end{equation}
where $\bm{R}$ denotes the lattice vector connecting the unit cells. The real-space overlap matrix elements $S_{i\alpha, j\beta}(\bm{R})$ are defined as:
\begin{equation}
    S_{i\alpha, j\beta}(\bm{R}_j-\bm{R}_i) = \int d^3 \bm{r} \phi^{*}_{i\alpha}(\bm{r}-\mathcal{R}_i) \phi_{j\beta}(\bm{r}-\mathcal{R}_j),
\end{equation}
where $\phi_{i\alpha}$ represents the $\alpha$-th orbital of the $i$-th atom located at $\mathcal{R}_i$ within the unit cell.

For non-periodic systems, Eq. (\ref{aoeig}) naturally reduces to the case where the unit cell is sufficiently large that no orbital overlaps with periodic images, and the $\bm k$-dependence vanishes.

\subsection{Neural Network Algorithm}
DeepH-pack predicts DFT Hamiltonians by leveraging a message-passing GNN where the atomic structure is mapped to graph data. In this representation, nodes correspond to atoms and edges connect atoms with finite Hamiltonian matrix elements. This GNN architecture naturally reflects the locality of the Hamiltonian and is consistent with the nearsightedness principle of electronic matter~\cite{prodan2005nearsightedness}. Through multiple layers of message-passing operations, the information from neighboring atoms is aggregated to update the node and edge features. This process effectively captures the complex local environment of the Hamiltonian matrix sub-blocks $H_{ij}$, allowing the model to predict accurate interactions even for atoms that are spatially separated but quantum-mechanically coupled.

A fundamental requirement for predicting physical quantities is to respect the symmetry of the system under global Euclidean transformations in 3-dimensional space (the \textit{E}(3) group). The Hamiltonian matrix block $H_{ij}$ is translationally invariant and rotationally equivariant~\cite{li2022deep,unke2021se}. To guarantee translational invariance, the network inputs rely solely on relative coordinate vectors $\mathcal{R}_{ij} = \mathcal{R}_j - \mathcal{R}_i$ rather than absolute positions. Regarding rotation, if the atomic coordinates are rotated by a rotation matrix $R\in$ \textit{SO}(3), where \textit{SO}(3) denotes the proper rotation group in 3-dimensional space, the Hamiltonian transforms according to 
\begin{equation}
    H_{ij}(\left\{ R \cdot \mathcal{R}_k \right\}) = D_i(R) H_{ij}(\left\{ \mathcal{R}_k \right\}) D^\dagger_j(R),
\end{equation}
where $D_i(R)$ and $D_j(R)$ are the representation matrices associated with the atomic-orbital basis sets centered at atoms $i$ and $j$, respectively, constructed as direct sums of Wigner $D$-matrices~\cite{li2022deep,wigner2012group}. To satisfy this constraint, DeepH-pack decomposes the target Hamiltonian into vectors that transform under irreducible representations (irreps) of the \textit{SO}(3) group ~\cite{gong2023general}. For systems involving SOC, the spin degrees of freedom introduce half-integer representations, but they can be transformed back into integer representations that fit into the framework of equivariant neural networks (e.g., $1/2 \otimes 1/2 = 0 \oplus 1$)~\cite{gong2023general}, ensuring that the predicted quantities strictly obey the physical transformation laws.

To achieve the symmetry constraints imposed on the outputs, DeepH-pack utilizes the equivariant neural network architecture~\cite{batzner2022e3}, and the internal features of the neural network are constructed as \textit{SO}(3)-equivariant vectors. These features carry specific irreps of the \textit{SO}(3) group and transform according to Wigner $D$-matrices. The core operation in the network involves the tensor product between these feature vectors and spherical harmonics $Y_{lm}(\hat{\mathcal{R}}_{ij})$, which encodes the geometric information of the bond directions. By stacking these equivariant operations, the network automatically preserves the complex symmetry patterns of the mapping from the geometric input to the Hamiltonian. Finally, the high-dimensional equivariant features are mapped back to the physical basis of the Hamiltonian via the inverse Clebsch-Gordan decomposition~\cite{gong2023general}, completing the prediction process.

DeepH-pack provides two distinct model architectures to balance accuracy and computational cost:  \texttt{Normal} and \texttt{Accurate}. The \texttt{Normal} model is a lightweight architecture based on improved DeepH-E3 ~\cite{gong2023general}, with optimizations for high inference speed and accuracy on standard tasks. For scenarios requiring higher precision or larger basis sets, the \texttt{Accurate} model offers enhanced expressivity. The computational complexity of standard equivariant tensor products (TP) in the \texttt{Normal} model typically scales as $O(L^6)$ with respect to the maximum order of angular momentum $L$. The \texttt{Accurate} model is built upon an improved DeepH-2 architecture~\cite{wang2024deeph} and integrates a speed-and-memory-optimized algorithm from the eSCN approach~\cite{passaro2023reducing}. This algorithm establishes a semi-local coordinate frame with the $z'$ axis aligned along the edge vector $\hat{\mathcal{R}}_{ij}$ while the $x'$ and $y'$ axes are not fixed. In this semi-local frame, the \textit{SO}(3) TP simplifies to an \textit{SO}(2) operation, reducing the complexity to $O(L^3)$. Building on this efficiency gain, the \texttt{Accurate} network can be scaled to larger and more complex architectures, paving the way for models with superior accuracy and generalizability. Furthermore, the \texttt{Accurate} model integrates optimized tensor product parameterization, attention mechanisms, and activation functions, achieving superior performance with moderate memory consumption.

\begin{figure}
    \centering
    \includegraphics[width=\linewidth]{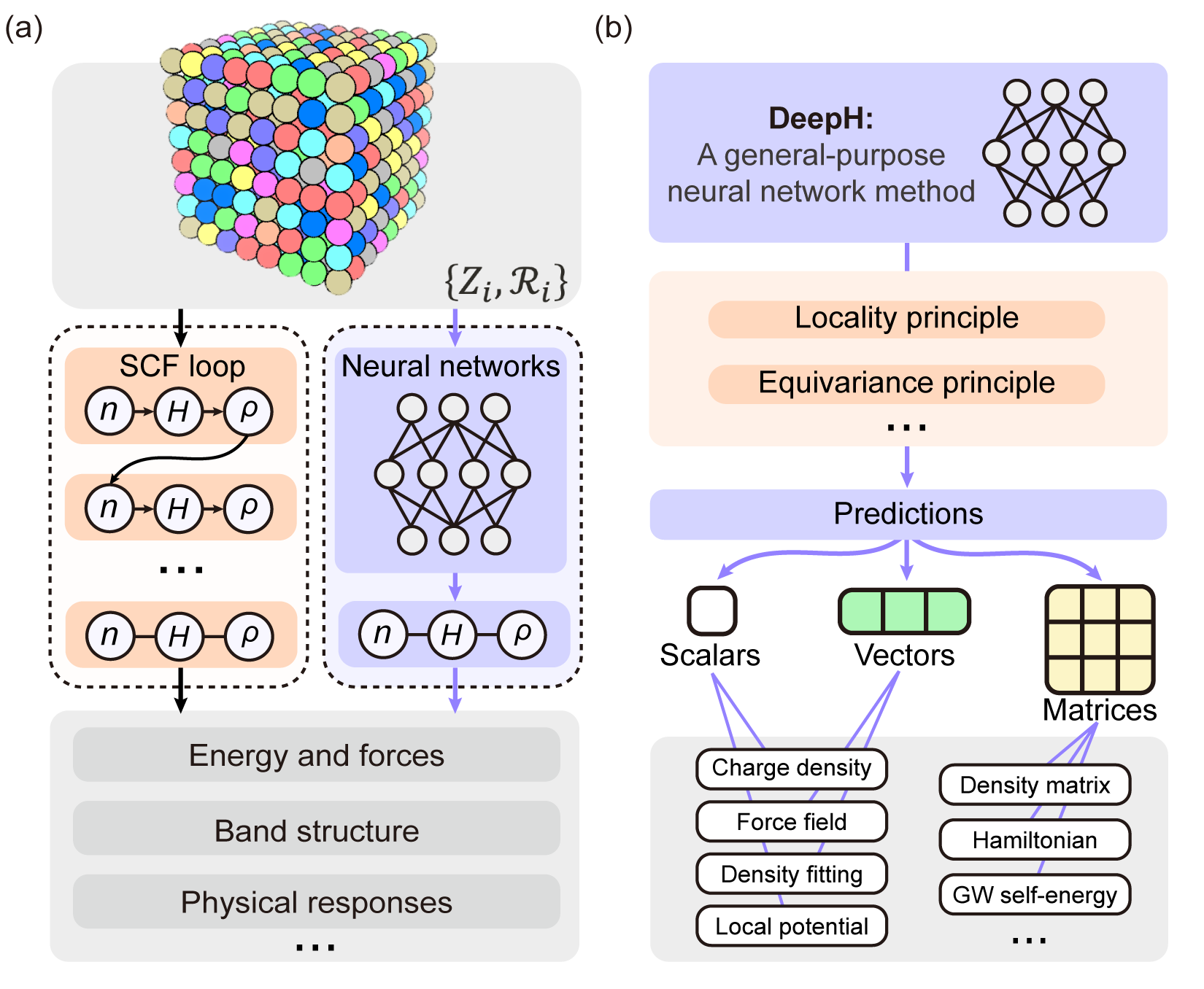}
    \caption{DeepH as a general-purpose neural network method. (a) Integration of DeepH with first-principles calculations. The DeepH model learns directly from converged density functional theory (DFT) data to predict electronic structures. These predictions serve a dual role: they enable efficient computation of material properties at first-principles accuracy, and provide high-quality initial guesses (such as charge density $n$, Hamiltonian $H$, and density matrix $\rho$) to accelerate DFT self-consistent field (SCF) convergence. (b) Physically constrained prediction of fundamental electronic structures. DeepH outputs scalar and tensor data that inherently satisfy physical constraints allowing systematic learning of diverse electronic properties. Its design ensures broad applicability across material systems and property types, establishing DeepH as a universal electronic structure algorithm.}
    \label{fig2}
\end{figure}

The architectural innovations described above establish DeepH as a general-purpose, efficient, and high-precision algorithmic framework (Fig. \ref{fig2}(b)). A key advantage is that the learning process is agnostic to the specific exchange-correlation functional employed in the first-principles calculations. Consequently, the model is capable of predicting DFT Hamiltonians derived from a wide spectrum of exchange-correlation functionals, provided that the training dataset maintains consistency. Furthermore, the applicability of DeepH extends beyond the typical DFT Hamiltonian. By adapting the output representations to match the specific tensor ranks, the framework can predict any physical quantity that transforms as an \textit{E}(3)-equivariant scalar or tensor. This versatility covers a diverse array of electronic structure properties, including electron-phonon coupling matrix elements, atomic forces and total energies, charge densities, density matrices, and quasiparticle self-energies within the $GW$ approximation.

\section{Workflow Support}
\begin{figure*}
    \centering
    \includegraphics[width=0.85\linewidth]{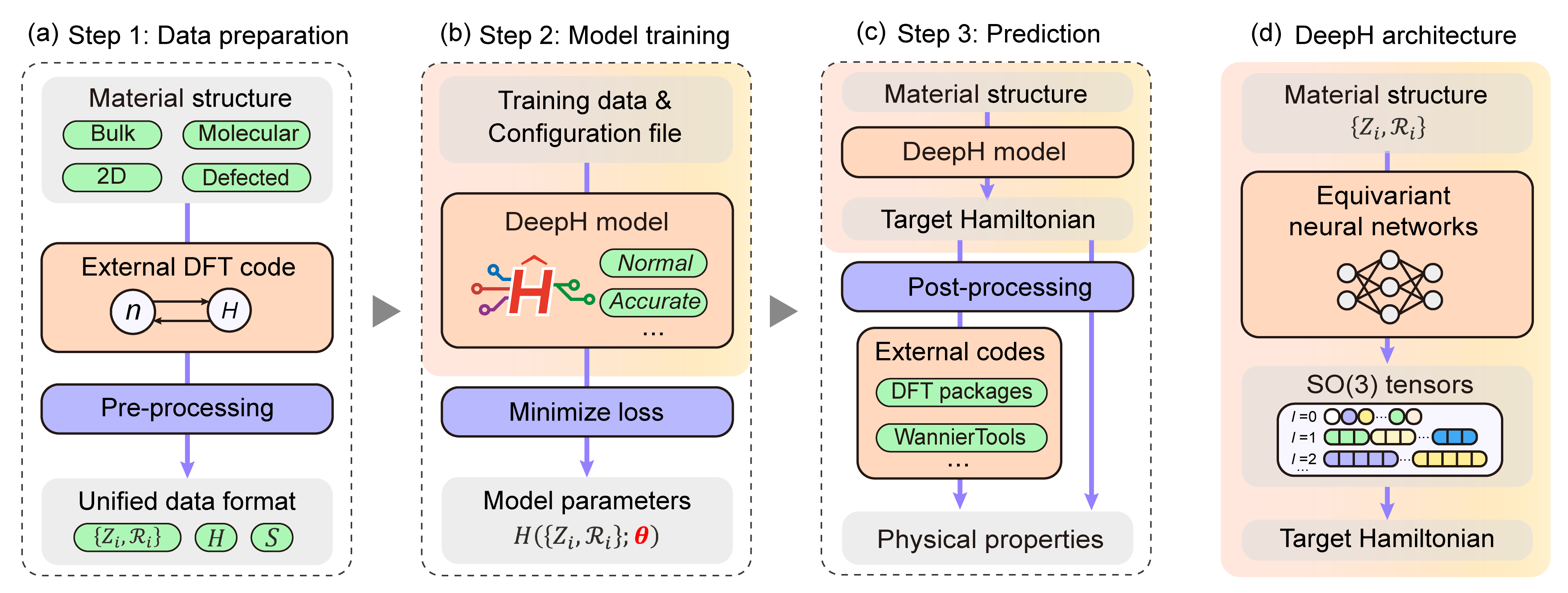}
    \caption{DeepH workflow schematic. The DeepH workflow comprises three core stages: (a) Data preparation: User-defined atomic structures are processed using DFT software to compute electronic structures. DeepH modules then construct graph representations encoding atomic positions (\(\{Z_i, \mathcal{R}_i\}\)), overlap matrices (\(S\)), and Hamiltonian labels (\(H\)). (b) Model training: DeepH models are trained with graph data using  suitable architectures and hyperparameters. The process minimizes loss functions to derive optimal parameters \(\theta\). (c) Prediction that includes model inference and Hamiltonian post-processing: Trained models predict Hamiltonians for new structures. DeepH-dock serves as an interface passing the Hamiltonian data to DFT software for continued calculations or post-processing tools for property analysis. (d) Internal mechanism of DeepH models: Atomic structures are encoded into \textit{SO}(3)-equivariant neural representations. These are transformed into Hamiltonians via the inverse Clebsch-Gordan decomposition.}
    \label{fig3}
\end{figure*}

As illustrated in Fig. \ref{fig3}, the typical research workflow within the DeepH framework proceeds sequentially from dataset generation and graph construction to model training, inference, and property post-processing. The platform supports a wide range of material systems, from bulk systems to complex defect structures, and offers multiple network architectures to balance accuracy and computational cost. To accommodate users across expertise levels, DeepH-pack implements a tiered interface framework progressing from accessibility to extensibility. First, a configuration-driven mode enables general users to deploy predefined architectures and training workflows through human-readable configuration files, positioning DeepH-pack as a versatile ecosystem for all-in-one computational materials solutions. Second, an object-oriented Python interface decomposes workflows into modular abstractions (data graphs, model blocks, training loops, etc.), facilitating the flexible assembly of custom architectures via class inheritance and composition. Finally, a low-level plugin system empowers developers to inject in-memory data across languages (Julia/C/Fortran) via MPI-based communication, enabling intervention in GPU training to modify intermediate network states (e.g., perturbing linear layer outputs) and seamless integration of external codes (e.g., DFT calculators), as demonstrated in the work of DeepH-Zero~\cite{li2024neural}. DeepH-pack utilizes the functional programming paradigm of JAX to implement efficient and differentiable electronic structure methods~\cite{bradbury2018jax}. By leveraging the Accelerated Linear Algebra (XLA) compiler, the software achieves hardware-agnostic execution across diverse computing architectures, including CPUs, GPUs, and TPUs. The framework incorporates optimized Just-In-Time (JIT) compilation strategies to handle variable array shapes during runtime, which results in significant reductions in memory usage during gradient backpropagation and substantial increases in computational throughput.

Complementing the core training package is DeepH-dock, a companion toolkit designed to manage the end-to-end workflow~\cite{deephdock2026li}. While DeepH-pack focuses on the computation-intensive tasks of model optimization and inference, DeepH-dock handles the essential tasks of data preparation and post-processing. It provides bidirectional conversion between the internal data standards and external DFT codes, including FHI-aims, Quantum ESPRESSO, SIESTA, ABACUS, and OpenMX, enabling users to seamlessly integrate neural network predictions into established research pipelines. Together, these two packages form a comprehensive system for dataset generation, Hamiltonian learning, and the subsequent calculation of physical properties.

\subsection{Pre-Processing Module}
\subsubsection{Input/Output Data Standards}
Data for training or testing the DeepH-pack are organized by individual material structures. To ensure efficient storage and high-throughput parallel input/output performance, we adopt a unified data format based on Hierarchical Data Format version 5 (HDF5). Specifically, the Hamiltonian and overlap matrices are stored using a sparse block representation, where each block corresponds to the matrix elements associated with a specific pair of atoms. This design naturally accommodates the mathematical operations required for \textit{SO}(3)-equivariant tensors. For a single data sample, the atomic coordinates and lattice vectors are recorded in a standard POSCAR file. The core electronic structure data, i.e., the overlap matrix and the DFT Hamiltonian, are stored in two separate files named $\texttt{overlap.h5}$ and $\texttt{hamiltonian.h5}$. These files share an identical sparse format. Inside the HDF5 structure, the non-zero matrix elements are flattened into a one-dimensional array named $\texttt{entries}$. To logically reconstruct the full matrices from $\texttt{entries}$, the sparse block structure is defined by an auxiliary dataset named $\texttt{atom\_pairs}$. Each row in $\texttt{atom\_pairs}$ consists of five integers $(i, j, k, n_1, n_2)$, representing a coupling between the $n_1$-th atom in the central unit cell and the $n_2$-th atom in the image cell specified by the lattice vector indices $(i, j, k)$. To facilitate the reconstruction of the matrices, the starting position and geometric dimension of each block in the $\texttt{entries}$ array are recorded in auxiliary datasets named $\texttt{chunk\_boundaries}$ and $\texttt{chunk\_shapes}$, respectively. The unit of energy for the Hamiltonian is electronvolts (eV). In addition to these matrix files, an $\texttt{info.json}$ file is required to provide essential metadata. This JSON file defines the global properties of the system, including the number of atoms, the angular momenta of atomic orbitals, spin polarization settings, and the Fermi level in eV. This standardized format allows DeepH-pack models to efficiently process complex Hamiltonian structures across different material systems.

\subsubsection{Structure Preparation}
In specialized applications, users often need to prepare custom datasets tailored to specific physical systems. For example, when modeling large supercells of amorphous materials, one may construct a training set from smaller representative amorphous structures. For moir\'e systems, a dataset can be built by generating untwisted layered van der Waals structures and then applying inter-layer shifts and perturbations. Sections 4 and 5 discuss such training-set design strategies in further detail.

For training universal materials models, atomic structures are commonly sourced from established materials databases such as the Materials Project~\cite{jain2013commentary}, ICSD~\cite{zagorac2019recent}, AFLOW~\cite{curtarolo2012aflow}, and Alexandria~\cite{schmidt2023machine}, which provide curated experimentally observed and computationally optimized configurations. Alternatively, traditional structure-search algorithms, including CALYPSO~\cite{wang2010crystal} and AIRSS~\cite{pickard2011ab} coupled with DFT software such as VASP~\cite{kresse1996efficiency}, can be employed to systematically explore high-quality atomic configurations. More recently, modern generative neural networks like MatterGen~\cite{zeni2025generative}, integrated with universal machine learning force fields, such as MatterSim~\cite{yang2024mattersim} and CHGNet~\cite{deng2023chgnet}, have enabled efficient and broad exploration of structural spaces. Together, these approaches offer practical and accessible pathways to obtaining diverse, high-quality, and stable atomic structures for training machine learning models.

\subsubsection{Data Conversion}
DeepH-dock supports the conversion of output files from a variety of DFT codes into our standard Hamiltonian data format. Upon executing the \texttt{dock\textvisiblespace{}convert} command, the software performs this conversion using parallelization across multiple cores on different compute nodes. 

It is also worth mentioning that, when given compatible atomic-orbital basis sets, DeepH-dock is able to project plane-wave (PW) DFT Hamiltonians onto atomic orbitals using the HPRO method proposed in ~\cite{gong2024generalizing}, thereby enabling training on PW data.

\subsubsection{Overlap Computation}
DeepH-dock incorporates a standalone module dedicated to the computation of overlap matrices, which serve as essential inputs for the DeepH-pack prediction workflow. This module implements the efficient grid-based interpolation scheme used in the HPRO package~\cite{gong2024generalizing}, which was originally proposed by Sankey and Niklewski~\cite{sankey1989ab}. By exploiting the finite support of numerical atomic orbitals as well as efficient spline interpolations, the algorithm achieves linear scaling with respect to the system size. In our benchmark testing, this workflow computes the overlap matrix for an 11,164-atom magic-angle twisted bilayer graphene system (using SIESTA's DZP basis set) within two minutes on a single CPU core.
Users may invoke overlap computation with the \texttt{dock\textvisiblespace{}compute\textvisiblespace{}overlap} command in DeepH-dock, or alternatively use overlap data generated by an external DFT code.

In addition to the built-in module, DeepH-dock supports the import of overlap matrices generated directly by compatible third-party DFT packages. For example, when using FHI-aims, users can obtain the required data by setting the control parameter $\texttt{sc\_iter\_limit}$ to zero. And when using ABACUS, users should set the option $\texttt{calculation}$ to $\texttt{get\_S}$. These configurations perform a single-step initialization without entering the self-consistency cycle, thereby extracting the overlap matrix with minimal computational overhead.

\subsection{Model Training and Inference}
\subsubsection{Training the Neural Network}
DeepH-pack is designed to handle diverse training scenarios, ranging from specialized datasets with a few hundred structures to large-scale materials databases containing hundreds of thousands of entries. To accommodate varying computational resources and accuracy requirements, the package provides two distinct neural network architectures named \texttt{Normal} and \texttt{Accurate}, which differ in parameter size and expressivity. A detailed description of the architectures is provided in Section 2. The training pipeline supports flexible data management, allowing users to either load datasets entirely into CPU memory for speed or stream them from disk to handle massive datasets. Furthermore, the implementation supports efficient multi-GPU parallelism across single or multiple computing nodes.

The training process begins by constructing a local coordinate graph representation of the crystal structure based on the sparsity pattern defined in $\texttt{overlap.h5}$ or $\texttt{hamiltonian.h5}$. In this graph, nodes represent individual atoms, while edges correspond to the non-zero Hamiltonian sub-blocks between atom pairs. Each node is assigned the atomic number of the corresponding atom, and each directed edge $(i,j)$ carries the relative position vector $\mathcal{R}_{ij}$ from atom $i$ to atom $j$. These graph features serve as the input to a JAX-based equivariant GNN, which preserves the physical symmetries of the underlying quantum interactions. For every connected pair of atoms $i$ and $j$, the network outputs the predicted Hamiltonian matrix sub-block ${H}_{ij}$.

The network parameters are optimized by minimizing a loss function $\mathcal{L}$ that quantifies the difference between the predicted Hamiltonian and the ground-truth labels stored in $\texttt{hamiltonian.h5}$. DeepH-pack allows users to select between Mean Squared Error (MSE) or Mean Absolute Error (MAE) as its optimization objective. Taking the MSE as an example, the loss for a batch of data is evaluated by:

\begin{equation}
    \mathcal{L} = \frac{1}{N_{b}} \sum_{k=1}^{N_{b}} \frac{1}{|\mathcal{N}_k|} \sum_{(i,j) \in \mathcal{N}_k} \frac{1}{N_{ij}}\left\| \hat{H}_{ij}^{(k)} - H_{ij}^{(k)} \right\|^2,
\end{equation}
where $N_b$ denotes the batch size, $|\mathcal{N}_k|$ denotes the number of $i$-$j$ pairs, $N_{ij}$ denotes the total number of matrix elements in the matrix sub-block ${H}_{ij}$,  $\mathcal{N}_k$ represents the set of interacting atom pairs in the $k$-th structure, and $\|\cdot\|$ denotes the Frobenius norm.

To streamline the workflow, the entire training procedure is encapsulated in a single command-line interface, $\texttt{deeph-train}$. The configuration for the training session is controlled by a TOML input file, which offers a user-friendly syntax compared to standard JSON formats. In this file, users specify critical hyperparameters, including the model architecture (e.g., \texttt{Normal} or \texttt{Accurate}), the learning rate schedule, the batch size, and the total number of training steps. This design ensures that the complex underlying operations of the JAX framework are accessible to users without requiring extensive programming knowledge.

\subsubsection{Auxiliary Functionalities for Hyperparameter Choice}
The training and inference workflows of DeepH are primarily controlled through a structured TOML configuration file. This file organizes all necessary settings into four core sections: The \texttt{system} section specifies the computational environment, including hardware declarations and resource allocation. The \texttt{data} section defines the dataset, including file paths, feature specifications, and metadata handling. The \texttt{model} section configures the neural network architecture, its components, the target physical quantities, and the associated loss function. The \texttt{process} section manages the execution workflow, encompassing training/inference procedures, convergence criteria, data loader settings, optimizers, and restart configurations.

Owing to space limitations, a full description of all TOML configuration options is not feasible within the scope of this article. The following outlines several core parameters critical for model setup; the complete documentation is available in the DeepH-pack Documentation~\cite{deephpack2026li}. Hardware resources are specified via \texttt{system.device} using the syntax \texttt{<type>*<num>:<id>}, which selects the type of the backend (e.g., \texttt{cpu}, \texttt{cuda}, \texttt{tpu}, \texttt{rocm}) and allocates computing devices. The neural network architecture is chosen by \texttt{model.net\_type}, which offers two primary presets in the standard version: the lightweight \texttt{Normal} network (typically with \(<\)1M parameters) is suitable for medium-scale Hamiltonian learning tasks, while the more advanced \texttt{Accurate} architecture (\(>\)5M parameters) is designed for high-accuracy modeling of complex systems. Another key parameter, \texttt{model.advanced.net\_irreps}, defines the irreducible representations of the network features in the form of an \texttt{e3nn.Irreps} string~\cite{geiger2022e3nn}, ensuring strict equivariance. For Hamiltonian prediction, the maximum angular momentum quantum number \(l_{\text{max}}\) specified must be at least twice the highest \(l\) value present in the Hamiltonian's basis set, in order to properly transform from a direct-product basis to a direct-sum basis. For example, if the system contains \(f\)-orbitals (\(l = 3\)), \(l_{\text{max}}\) must be no less than 6. When the SOC is turned on, \(l_{\text{max}}\) needs to be further increased by 1 to include the spin degree of freedom.

\subsubsection{Predicting DFT Hamiltonians}
After the training phase is complete, the optimized model can be deployed to predict DFT Hamiltonians for unseen structures. DeepH-pack is designed to support high-throughput inference tasks and efficiently utilizes multi-GPU acceleration to process large batches of materials. The prediction workflow is managed via a TOML configuration file and is initiated by the single-line command $\texttt{deeph-infer}$.

For a specific target material, the inference engine requires three essential inputs: the atomic structure defined in a $\texttt{POSCAR}$ file, the metadata in $\texttt{info.json}$, and the precomputed overlap matrix. These input files can be conveniently prepared using the DeepH-dock interface. The final output is the predicted Hamiltonian matrix stored in the standardized sparse HDF5 format described in section 3.1.1 and named \texttt{hamiltonian\_pred.h5}.

\subsection{Post-Processing Module}
\subsubsection{Band Structure and Density of States}
\begin{figure}
    \centering
    \includegraphics[width=0.8\linewidth]{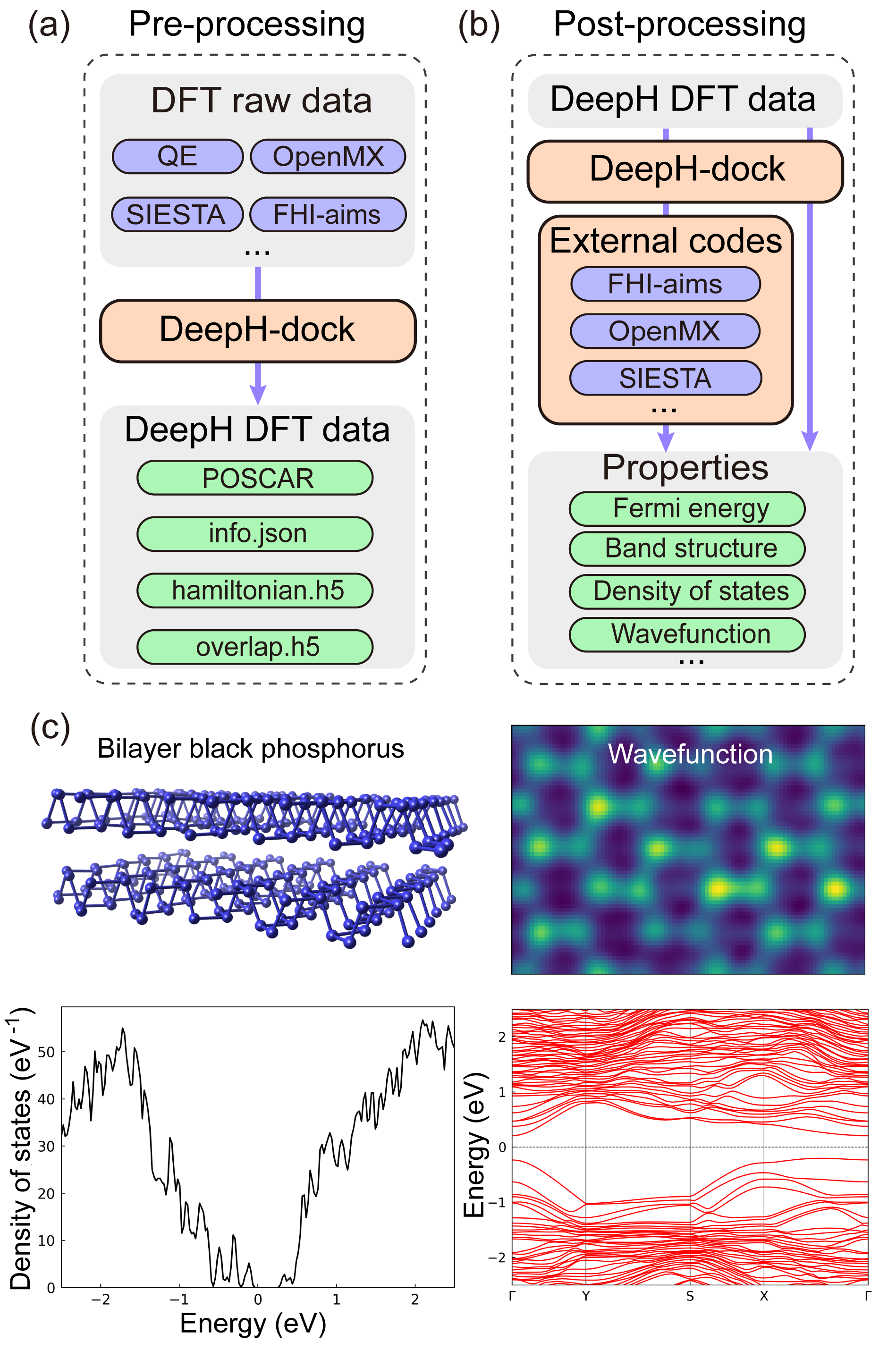}
    \caption{Schematic illustration of DeepH-dock functionality. (a) Serving as the interface for DeepH-pack, DeepH-dock converts data from multiple popular DFT software packages into a unified format compatible with DeepH-pack.(b) DeepH-format data, either predicted by DeepH-pack or converted from other DFT software, can be further processed to compute target material properties. (c) DeepH-pack predicts Hamiltonians, enabling DeepH-dock to directly output key electronic properties (DOS, wavefunction visualization for the valence band maximum at $\Gamma$ point, and band structure) for perturbed bilayer phosphorene.}
    \label{fig4}
\end{figure}

The electronic band structure $\mathcal{E}_n(\bm{k})$ defined in Eq.(\ref{aoeig}) describes the single-particle energy levels of electrons in periodic crystals, and is one of the most important physical quantities that can be extracted from the DFT Hamiltonian. For systems of moderate sizes ($<1000$), DeepH-dock can directly solve the generalized eigenvalue problem in Eq. (\ref{aoeig}) and thus produce the band structure from the Hamiltonian predicted by DeepH-pack. In addition, DeepH-dock can determine the Fermi energy according to the electronic occupation number and compute the electronic DOS from the energy levels. Figure \ref{fig4} shows the band structure and DOS of bilayer black phosphorus obtained in a post-processing step from a Hamiltonian predicted by DeepH-pack.

\subsubsection{Wave Functions}
Wave functions are another key quantity DFT calculations can provide. They describe the spatial distribution of electrons for each energy level, and are required for many subsequent physical-property calculations, such as electronic conductivity~\cite{scheidemantel2003transport}, optical spectrum~\cite{gajdovs2006linear}, and many-body perturbation theory corrections~\cite{hybertsen1986electron}. The Bloch wave function is defined by
\begin{equation}
    \psi_{n \bm{k}}(\bm{r}) = \frac{1}{\sqrt{N}}\sum_{\bm{R}}\sum_{j\beta \in \text{uc}} e^{i\bm{k}\cdot \bm{R}} c_{n, j\beta}(\bm{k}) \phi_{j\beta}(\bm{r}-\bm{R}-\mathcal{R}_j),
\end{equation}
where $\bm{R}$ denotes the lattice translation vector, and $N$ is the size of the Born-von Karman supercell. Note that only the atomic positions $\mathcal{R}_j$ within the unit cell are considered here.

After solving Eq. (\ref{aoeig}), DeepH-dock can export the wave function coefficients $c_{n, j\beta}(\bm{k})$ to a HDF5 file. Furthermore, DeepH-dock can compute and output the periodic part of the Bloch wave function in real space via 
\begin{equation}
    u_{n \bm{k}}(\bm{r}) = \frac{1}{\sqrt{N}}\sum_{\bm{R}}\sum_{j\beta \in \text{uc}} e^{-i\bm{k}\cdot (\bm{r}-\bm{R})} c_{n, j\beta}(\bm{k}) \phi_{j\beta}(\bm{r}-\bm{R}-\mathcal{R}_j).
\end{equation}

\subsubsection{Charge Density and Density Matrices}
DeepH-dock can compute the one-particle density matrix under the atomic orbital basis, defined as
\begin{equation}
\label{dm}
\rho_{i\alpha,j\beta}\left(\bm{k} \right) = \sum_{n} f_{n\bm{k}} c_{n,i\alpha}(\bm{k}) c^{*}_{n,j\beta}(\bm{k}) ,
\end{equation}
where $i,j$ index atoms in the unit cell, $\alpha,\beta$ label orbitals on each atom, $f_{n\bm{k}}$ denotes the Fermi-Dirac occupation factor of state $(n,\bm{k})$, and the summation runs over all bands. DeepH-dock offers two algorithms for its construction: (i) a direct evaluation according to Eq. (\ref{dm}) when wave function coefficients are available, and (ii) a diagonalization-free approach based on the PEXSI method~\cite{lin2014siesta}, which computes the density matrix without explicitly solving for the full band-structure eigenvalue problem and exhibits sub-cubic scaling with the number of atoms.

\subsubsection{Interfacing with DFT software}
DeepH-dock enables users to export the predicted data, including Hamiltonian matrices, overlap matrices, and density matrices, in formats compatible with a variety of DFT codes. Currently supported interfaces include:

1) SIESTA Interface: The \texttt{*.HSX} Hamiltonian/overlap matrix file and \texttt{*.DM} density-matrix file. The \texttt{*.HSX} file stores the Hamiltonian and overlap matrices in a compact binary format. This file can be used for band structure calculations and band unfolding analysis, enabling the study of electronic properties in supercells and interfaces. The \texttt{*.DM} file contains the density matrix, which is essential for restarting SCF calculations and for post-processing analyses such as charge density and electronic population analysis.

2) Quantum ESPRESSO Interface: The charge-density HDF5 file. This \texttt{*.h5} format file stores the charge density output from Quantum ESPRESSO calculations. It serves as a critical input for restarting SCF calculations and can be utilized for subsequent electronic property computations, including electrostatic potential analysis and charge transfer studies.

3) FHI‑aims interface: The \texttt{hamiltonian.out} and \texttt{overlap-matrix.out} file of FHI‑aims. These plain-text files contain the Hamiltonian and overlap matrix data, respectively. Currently, these files are suitable for cluster calculations at the gamma point only. For periodic crystal properties and predictions, an interface utilizing the FHI-aims Atomic Simulation Interface (ASI)~\cite{stishenko2023atomic} is required, which is currently in the beta testing phase in DeepH-dock.

4) OpenMX Interface: The \texttt{*.scfout} file that contains Hamiltonian, overlap matrix, density matrix, and position matrix. The \texttt{*.scfout} file is a comprehensive output that includes Hamiltonian, overlap matrix, density matrix, and position matrix data. This file format supports a wide range of post-processing tasks, such as band structure calculations, topological property analysis (e.g., Berry curvature and Chern number), optical/response properties calculations, and other advanced electronic structure computations.

5) ABACUS Interface: The \texttt{*.csr} sparse format files which include Hamiltonian, overlap matrix, density matrix, and position matrix. These files store the Hamiltonian, overlap matrix, density matrix, and position matrix in compressed sparse row (CSR) format. The CSR format enables efficient storage and manipulation of sparse matrices, facilitating band structure calculations, subsequent topological property analyses, and optical/response properties calculations.

By leveraging these interfaces, users can load predictions from DeepH-pack into their preferred DFT workflow and thereby explore a wider range of physical properties and material phenomena.

\section{Benchmarking}
To rigorously evaluate the accuracy and efficiency of DeepH-pack, we employ a comprehensive set of benchmarks spanning diverse dimensionalities and chemical environments. The test systems range from molecular systems (ethanol) to two-dimensional (2D) materials (monolayer $\text{MoS}_2$, bilayer $\text{Bi}_2\text{Se}_3$) and bulk crystals (graphite, $\text{SiC}$, and carbon allotropes). Crucially, these datasets are selected to address critical physical challenges, including relativistic SOC effects in $\text{Bi}_2\text{Se}_3$, defect states in $\text{SiC}$, and structural polymorphism in carbon. The performance of the networks implemented in DeepH-pack, namely the \texttt{Normal} and \texttt{Accurate} architectures, is evaluated, including training/inference time and memory usage. A comparison with the previous DeepH-E3 implementation is also provided. All training and inference processes were performed on NVIDIA RTX 4090 GPUs with AMD EPYC 7502 Processors. The model accuracy is assessed based on a set of standardized metrics listed below. The error between the predicted Hamiltonian $H_{i\alpha,j\beta}$ and the benchmark result is evaluated as the most direct measurement of model accuracy. The accuracy of observable quantities is also evaluated, including the mean absolute error (MAE) in band energies ($\mathcal{E}_{n(\bm{k})}$), the MAE in DOS ($D_{(\varepsilon)}$), and the root mean square error (RMSE) of eigenstates ($\psi_{n(\bm{k})}$) at $\Gamma$ point. The metrics are defined below.

\begin{equation}
\text{MAE}_H=\frac{\sum_{i\alpha,j\beta} |H_{i\alpha,j\beta}^\text{pred}-H_{i\alpha,j\beta}^\text{bm}-\Delta\mu_H S_{i\alpha,j\beta}|}{N_H},
\end{equation}

\begin{equation}
\text{MAE}_\text{band}=\frac{\int d^n\bm{k} \sum^{\prime}_n |\mathcal{E}^\text{pred}_{n(\bm{k})}-\mathcal{E}^\text{bm}_{n(\bm{k})}-\Delta\mu_\text{band}|}{\int d^n\bm{k} \ N^{\prime}_{\text{band}(\bm{k})}},
\end{equation}

\begin{equation}
\text{MAE}_\text{DOS}=\frac{\int_{\varepsilon_\text{min}}^{\varepsilon_\text{max}} |D^\text{pred}_{(\varepsilon+\Delta\mu_\text{band})}-D^\text{bm}_{(\varepsilon)}|d\varepsilon}{\int_{\varepsilon_\text{min}}^{\varepsilon_\text{max}} D^\text{bm}_{(\varepsilon)}d\varepsilon},
\end{equation}

\begin{equation}
\text{RMSE}_{\psi,\bm{k}=\bm{0}}=\left(\frac{\sum^{\prime}_n (1-\sum^{\prime\prime}_{g} |\langle \psi^\text{bm}_{g(\bm{k}=\bm{0})}|\psi^\text{pred}_{n(\bm{k}=\bm{0})} \rangle|^2)}{N^{\prime}_{\text{band}(\bm{k}=\bm{0})}}\right)^{1/2}.
\end{equation}

The reciprocal space is sampled sufficiently to evaluate the energy bands and DOS. For ethanol, only the $\Gamma$ point is included. For periodic crystals, the Brillouin zones are sampled with a relatively dense grid with a $k$-point spacing of 0.02--0.07$\text{\AA}^{-1}$. Gaussian smearing is applied to evaluate DOS with a smearing width of 0.01$\sim$0.08 eV. To filter out information near the Fermi level, the errors of the energy bands, DOS, and eigenstates are calculated within an energy window $(\varepsilon_\text{min},\ \varepsilon_\text{max})$. This is indicated by $\sum^{\prime}$, which means $\sum^{\prime}_n=\sum_{n, \ s.t. \ \varepsilon_\text{min}<\mathcal{E}_{n(\bm{k})}^\text{bm}<\varepsilon_\text{max}}$. The energy window is initially centered at the Fermi energy $\varepsilon_F$ with a total width of 6.0 eV, and is expanded until at least 10 energy bands are included (except for water molecules, where the 4 occupied and the 7 lowest unoccupied energy levels are included). In the denominator, $N_H$ is the total number of elements in the Hamiltonian matrix, and $N^{\prime}_{\text{band}(\bm{k})}$ is the number of energy levels within the energy window. The note $\sum^{\prime\prime}$ in the expression of $\text{RMSE}_\psi$ denotes the summation restricted within the nearly degenerate subspace related to $\psi_n$, which is identified by an energy spacing threshold $\varepsilon_\text{th}$ (defaulting to 5 meV). The inner product between eigenstates is calculated as $\langle \psi_{m(\bm{k})}|\psi_{n(\bm{k})}\rangle=c_{m(\bm{k})}^\dagger S_{(\bm{k})}c_{n(\bm{k})}$, where $S_{(\bm{k})}$ is the Fourier transform of the overlap matrix and $c_{n(\bm{k})}$ is the eigenvector from diagonalization. The errors of the Hamiltonian matrix, band energies, and DOS are defined after gauge correction $\Delta\mu$~\cite{wang2024deeph}.

\begin{equation}
\Delta\mu_H=\frac{\sum_{i\alpha,j\beta} (H^\text{pred}_{i\alpha,j\beta}-H^\text{bm}_{i\alpha,j\beta}) S_{i\alpha,j\beta}}{\sum_{i\alpha,j\beta} S_{i\alpha,j\beta}^2}
\end{equation}

\begin{equation}
\Delta\mu_\text{band}=\text{median}\{\mathcal{E}^\text{pred}_{n(\bm{k})}-\mathcal{E}^\text{bm}_{n(\bm{k})}\ |\ \varepsilon_\text{min} < \mathcal{E}^\text{bm}_{n(\bm{k})}<\varepsilon_\text{max}\}
\end{equation}
The results are shown in Table \ref{table1}, and the best-performing values for each metric are highlighted in bold.

\begin{table*}[h!]
\caption{The efficiency and accuracy of DeepH-pack on benchmark datasets. The bold items mark the best result across tested models.}
\begin{adjustbox}{center}
\begin{tabular}{crrrrrrr}
\hline
\addlinespace[0.5ex]
~ & ~ & \makecell[r]{Water\\molecule} & \makecell[r]{Monolayer\\$\text{MoS}_2$} & \makecell[r]{Bilayer\\$\text{Bi}_2\text{Se}_3$} & Graphite & \makecell[r]{Defective\\$\text{SiC}$} & \makecell[r]{Carbon\\allotropes} \\
\addlinespace[0.5ex]
\hline
\addlinespace[0.5ex]
\makecell{Number of atoms\\ per unit cell} & ~ & 3 & 75 & 90  & 108 & 64  & 3$\sim$58 \\
\addlinespace[0.5ex]
\hline
\addlinespace[0.5ex]
Training set size & ~ & 3000 & 300 & 460 & 240 & 300 & 342 \\
Batch size & ~ & 30   & 1   & 1   & 1   & 1   & 1   \\
\addlinespace[0.5ex]
\hline
\addlinespace[0.5ex]
\multirow{3}{*}{\makecell{Hamiltonian matrix\\MAE (meV) $\downarrow$}} & DeepH-E3~\cite{gong2023general} & 0.25 & 0.43 & 0.27 & 0.36 & 0.20 & 1.8 \\[1ex]
~ & Normal & 0.11 & 0.13 & 0.077 & 0.21 & 0.058 & 0.69 \\
~ & Accurate   & \textbf{0.077} & \textbf{0.041} & \textbf{0.044} & \textbf{0.13} & \textbf{0.037} & \textbf{0.53} \\
\addlinespace[0.5ex]
\hline
\addlinespace[0.5ex]
\multirow{3}{*}{\makecell{Band energy MAE\\(meV) $\downarrow$}} & DeepH-E3 & 2.4 & 0.74 & 0.97 & 12 & 0.50 & 45 \\[1ex]
~ & Normal & 1.3 & 0.22 & \textbf{0.29} & 12 & 0.24 & 17 \\
~ & Accurate   & \textbf{0.71} & \textbf{0.060} & 0.31 & \textbf{8.7} & \textbf{0.20} & \textbf{10} \\
\addlinespace[0.5ex]
\hline
\addlinespace[0.5ex]
\multirow{3}{*}{\makecell{Density of states MAE\\ ($\times10^{-3}$, dimensionless) $\downarrow$}} & DeepH-E3 & -- & 8.5 & 1.8 & 17 & 0.61 & 27 \\[1ex]
~ & Normal & -- & 2.4 & \textbf{0.65} & 16 & 0.27 & 17 \\
~ & Accurate   & -- & \textbf{0.70} & 0.70 & \textbf{12} & \textbf{0.21} & \textbf{11} \\
\addlinespace[0.5ex]
\hline
\addlinespace[0.5ex]
\multirow{3}{*}{\makecell{Eigenstate RMSE\\($\times10^{-2}$, dimensionless) $\downarrow$}} & DeepH-E3 & 0.22 & 7.2 & 5.1 & 7.0 & 2.0 & 11 \\[1ex]
~ & Normal & 0.13 & 1.8 & \textbf{2.9} & 7.9 & 0.76 & 5.2 \\
~ & Accurate   & \textbf{0.12} & \textbf{0.61} & 3.5 & \textbf{3.9} & \textbf{0.70} & \textbf{2.2} \\
\addlinespace[0.5ex]
\hline
\addlinespace[0.5ex]
\multirow{3}{*}{\makecell{Training time\\(s/batch) $\downarrow$}} & DeepH-E3 & 0.63 & 1.0 & 1.3 & 0.63 & 1.1 & 0.92 \\[1ex]
~ & Normal & \textbf{0.052} & \textbf{0.12} & \textbf{0.23} & \textbf{0.14} & \textbf{0.21} & 0.23 \\
~ & Accurate   & 0.076 & 0.19 & \textbf{0.23} & 0.23 & 0.31 & \textbf{0.17} \\
\addlinespace[0.5ex]
\hline
\addlinespace[0.5ex]
\multirow{3}{*}{\makecell{Inference time\\(s/structure) $\downarrow$}} & DeepH-E3 & 0.12 & 0.17 & 0.24 & 0.11 & 0.21 & 0.19 \\[1ex]
~ & Normal & \textbf{0.02} & \textbf{0.03} & \textbf{0.05} & \textbf{0.04} & \textbf{0.06} & \textbf{0.06} \\
~ & Accurate   & 0.03 & 0.07 & 0.13 & 0.13 & 0.14 & 0.08 \\
\addlinespace[0.5ex]
\hline
\addlinespace[0.5ex]
\multirow{2}{*}{\makecell{Inference memory\\(MB/atom) $\downarrow$}} & Normal & \textbf{0.9} & 7.9 & 13 & 6.5 & 13 & 22 \\
~ & Accurate   & \textbf{0.9} & \textbf{6.4} & \textbf{9.1} & \textbf{5.9} & \textbf{12} & \textbf{7.9} \\
\addlinespace[0.5ex]
\hline
\addlinespace[0.5ex]
\multirow{3}{*}{\makecell{Parameter size\\(million)}} & DeepH-E3 & 0.68 & 0.77 & 1.46 & 0.42 & 0.72 & 0.84 \\[1ex]
~ & Normal & 1.36 & 2.06 & 1.29 & 0.57 & 1.98 & 3.48 \\
~ & Accurate & 3.58 & 8.39 & 7.52 & 3.64 & 8.39 & 2.62 \\
\addlinespace[0.5ex]
\hline
\end{tabular}
\end{adjustbox}
\label{table1}
\end{table*}

We select water as a representative of small molecules. The dataset contains 4999 configurations of water molecules from the MD17 dataset~\cite{chmiela2017machine}, which consists of structures sampled from molecular dynamics trajectories. The DFT calculations are performed by AI2DFT~\cite{li2024neural} with norm-conserving (NC) pseudopotentials~\cite{hamann2013optimized} and a double-zeta polarized (DZP) basis set~\cite{junquera2001numerical}. The generalized gradient approximation (GGA) for the exchange-correlation functional in the form of Perdew-Burke-Ernzerhof (PBE) is used~\cite{perdew1996generalized}. The \texttt{Normal} and \texttt{Accurate} models converged after 8 hours of training, with their performance evaluated on the same test set. Both \texttt{Normal} and \texttt{Accurate} models outperform DeepH-E3, achieving inference speeds 3 to 5 times faster and reducing errors by 50$\sim$70$\%$. The \texttt{Accurate} model predicts the most accurate Hamiltonian matrix and energy levels, while the \texttt{Normal} model achieves a compatible accuracy with much fewer parameters and less inference time. Only the 4 occupied and 7 lowest unoccupied states are included in the energy MAE and eigenstate RMSE analysis, as higher unoccupied states are not practically relevant to most of the property calculations and experiments. It should be noted that the software and calculation setup in this work are different from the published MD17 dataset, and the Hamiltonian matrix error cannot be directly compared with other works~\cite{unke2021se,yu2023efficient}. 

Monolayer $\text{MoS}_2$ and bilayer $\text{Bi}_2\text{Se}_3$ were selected as representative 2D materials. Both are van der Waals materials and can be grown into thin layers~\cite{novoselov20162d,liu2016van}. DFT calculations are performed by OpenMX~\cite{ozaki2003variationally} with norm-conserving (NC) pseudopotentials~\cite{morrison1993nonlocal} and a multiple-zeta polarized basis set~\cite{ozaki2004numerical}. These datasets have been used in the work of DeepH-E3~\cite{gong2023general}. $\text{MoS}_2$ belongs to the transition metal dichalcogenide (TMD) family and shows outstanding electronic and optical qualities, notably high carrier mobility and strong excitonic effects~\cite{xiao2012coupled,ugeda2014giant,liu2021promises}. These properties enable its potential applications in transistors and photodetectors. The dataset is constructed by $5\times5$ supercells with random perturbation of atom positions, resulting in a total of 500 structures. DeepH-pack demonstrates robust performance on these datasets. The \texttt{Normal} network achieves significantly higher accuracy (reducing MAE by $\sim70\%$) than DeepH-E3 with the smallest time cost in training and inference processes. The \texttt{Accurate} network further improves the accuracy by approximately 3-fold with lower memory consumption than \texttt{Normal} network. The four accuracy metrics show the same trend that \texttt{Accurate} network performs the best, \texttt{Normal} network in the middle, both better than DeepH-E3. $\text{Bi}_2\text{Se}_3$ is known by its high thermoelectric efficiency and non-trivial topological properties~\cite{zhang2009topological}, and is a typical material with significant SOC effects. The dataset contains 576 structures, each one is a $3\times3$ supercell, applying in-plane interlayer shifts between the two layers, and random perturbing the atoms in each direction by $0.1\AA$. SOC was included in the OpenMX calculations because it plays an essential role in the band inversion and leads to the non-trivial band topology. To handle the additional spin angular momentum of $1/2 \otimes 1/2=0 \oplus 1$, the maximum angular momentum in the network needs to be 1 plus twice the maximum angular quantum number of the basis set orbitals, that is, $l_\text{max}=2\times2+1=5$ in this case. \texttt{Normal} and \texttt{Accurate} networks demonstrate comparable overall performance on this dataset. Specifically, \texttt{Accurate} network achieves smaller Hamiltonian MAE, while \texttt{Normal} network yields slightly smaller error in energy bands, DOS, and eigenstates. Even for the heavy \texttt{Accurate} model with 7$\sim$8 million parameters, the training on the two datasets can be completed within 48 hours, which greatly facilitates the hyperparameter optimization and accelerates the research works.

Graphite, $\text{SiC}$, and carbon allotropes were chosen to represent bulk systems. The training data for graphite and carbon allotropes are calculated by OpenMX, and the training data for $\text{SiC}$ is calculated by SIESTA~\cite{soler2002siesta} with norm-conserving (NC) pseudopotentials~\cite{hamann2013optimized} and a double-zeta polarized (DZP) basis set~\cite{junquera2001numerical}. Graphite has a layered structure analogous to the $\text{MoS}_2$ mentioned above, and is representative of 3D van der Waals materials. Furthermore, a variety of stacking modes (e.g. AA, AB, ABC) dictate diverse mechanical properties and electronic structures, for instance, the strong correlation effects in rhombohedral graphite~\cite{hagymasi2022observation,han2024correlated}. The dataset is constructed by $3\times3\times6$ cell expansion of the AB stacked graphite. Random perturbations of atom positions are then added to give 300 distinct structures. Silicon carbide ($\text{SiC}$), a semiconductor possessing wide band gap and large thermal conductance, is chosen as a representative of materials with defects~\cite{yuan2024deep}. To address the challenge posed by the intrinsic point defects on device performance, a dataset of 500 structures is generated to model the anti-site defects. Each structure is created by random replacement between carbon and silicon atoms in a $2\times2\times2$ supercell. The relative performance between the networks on the graphite and $\text{SiC}$ dataset remains consistent with prior observations on the $\text{MoS}_2$ dataset, whereas \texttt{Normal} network outperforms in time cost and \texttt{Accurate} network outperforms in model accuracy. Note that the memory cost per atom for the $\text{SiC}$ system is larger than the $\text{MoS}_2$ dataset, because the number of neighbors in bulk materials ($\sim$130 for the $\text{SiC}$ dataset) is larger than that in 2D materials ($\sim$60 for monolayer $\text{MoS}_2$). The last dataset contains 427 carbon allotropes from the Samara Carbon Allotrope Database~\cite{hoffmann2016homo}, all of which are stable structures. The dataset sparsely samples a large configurational space with a limited number of data points, which poses a significant challenge for model generalization. In this complicated case study, \texttt{Accurate} network exhibits expressivity and generalization capabilities superior to those of the other networks. The \texttt{Accurate} network achieves a remarkably low band MAE of 10 meV and relative errors of approximately 0.01 for the DOS and 0.02 for the eigenstates. These high-accuracy results strongly support the reliability of subsequent property calculations (e.g., optical and response properties) derived from the predicted Hamiltonian. Training on the graphite dataset requires 60 hours to reach full convergence. However, by applying a smaller patience in the learning rate scheduler, the time cost can be considerably reduced with an accuracy drop of less than 10$\%$. For the $\text{SiC}$ and carbon allotropes datasets, the training processes are completed within 48 hours, which is completely acceptable.

Benefiting from the optimized code engineering in \texttt{e3nn\_jax}~\cite{geiger2022e3nn} and DeepH-pack, the \texttt{Normal} network, which has the same architecture as DeepH-E3, outperforms DeepH-E3, reducing errors by $\sim40\%$ to $\sim70\%$ and increasing efficiency by 100\% to 800\%. The \texttt{Accurate} network achieves even $\sim20\%$ to $\sim40\%$ lower error due to advanced architecture design and larger parameter size, especially on general-purpose datasets such as the carbon allotropes dataset. Designed for large-capacity general-purpose material modeling, the \texttt{Accurate} model achieves greater memory efficiency than the \texttt{Normal} model through its use of a semi-local coordinate frame, thus enabling larger-scale training.

\section{Example Applications}
To demonstrate the practical utility and workflow versatility of DeepH-pack, we present three representative applications. These examples illustrate the framework's adaptability across different software ecosystems and its capacity to address diverse computational challenges in physics and materials research, including tractable first-principles calculations for large systems, high-precision characterization of low-energy physics, efficient modeling of complex local environments, and the establishment of generalizable foundation models.

\subsection{Twisted bilayer MoTe\(_2\)}
Twisted bilayer TMDs have emerged as a prominent research focus in condensed matter physics and materials science~\cite{jin2019observation,wang2020correlated,ghiotto2021quantum,park2023observation}. By precisely controlling the interlayer twist angle, their electronic structures can be effectively engineered, giving rise to a rich variety of correlated electronic states and topological properties. Among them, twisted bilayer MoTe\(_2\) has attracted significant attention due to the potential emergence of correlated insulating states, topologically non-trivial bands, and possible superconducting behavior near specific ``magic'' angles, making it an ideal platform for exploring 2D strongly correlated physics and topological quantum phenomena. However, performing full first-principles calculations on such systems is highly challenging because the size of the moir\'e superlattice increases dramatically as the twist angle decreases. The computational cost of conventional DFT scales cubically with the number of atoms, rendering systematic angle-dependent scanning computationally prohibitive. Therefore, developing approaches that retain first-principles accuracy while achieving high computational efficiency is particularly urgent. This creates an important application scenario for machine learning methods, especially the DeepH framework, which can efficiently predict the DFT Hamiltonians directly from atomic structures. By constructing a highly transferable Hamiltonian model, DeepH holds the potential to enable high-throughput computation and rapid screening across multiple degrees of freedom such as twist angle and stacking order, thereby systematically revealing the structure-property relationships in moir\'e materials.

\begin{figure}
    \centering
    \includegraphics[width=\linewidth]{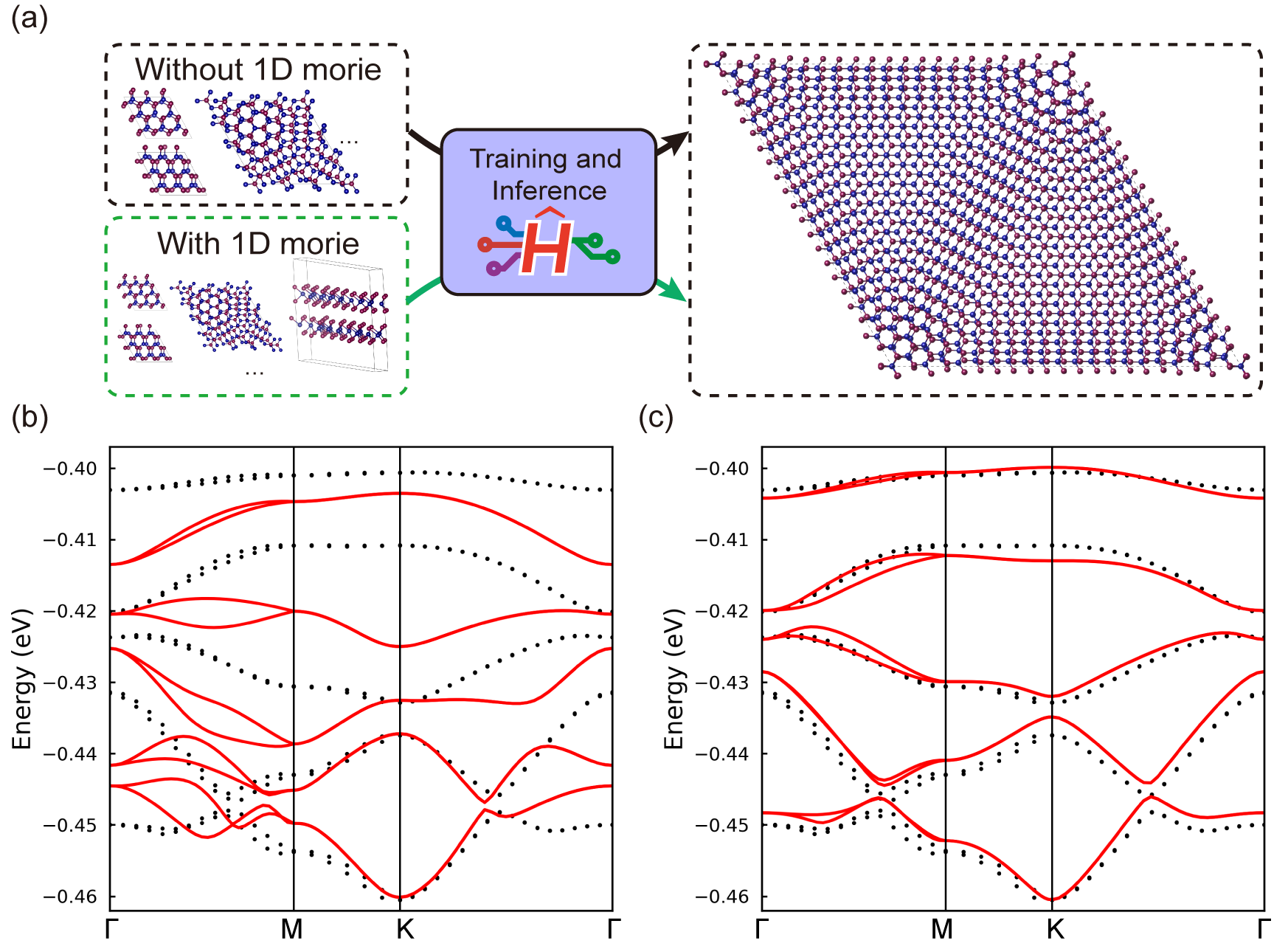}
    \caption{Influence of training‑set composition on DeepH's prediction accuracy for a large‑scale twisted bilayer MoTe$_2$ system. (a) Schematic diagram of the two training‑set generation strategies. Strategy I includes interlayer‑shifted bilayer and twisted bilayer structures. Strategy II adds a variety of uniquely designed quasi-1D moir\'e structures. (b) Band structure of a 1,986-atom twisted bilayer MoTe$_2$ predicted by the DeepH model trained using Strategy I (red solid lines), compared with the DFT reference results (black dotted lines). The mean absolute error (MAE) of the band energies is 4.64 meV. (c) Band structure of the same system predicted by the model trained with Strategy II (red solid lines), showing improved agreement with the DFT reference (black dotted lines). The MAE is reduced to 1.26 meV.}
    \label{fig5}
\end{figure}

High-accuracy prediction of Hamiltonians for supercell structures with DeepH typically begins by training a specialized model on a set of smaller-scale configurations. Since DeepH directly predicts Hamiltonian matrix elements between atom pairs, knowledge of local interactions learned from smaller systems can be effectively transferred to much larger structures, enabling rapid electronic-structure prediction at DFT accuracy. The selection of training structures is crucial when building a specialized model. The guiding principle is that the bonding environments present in the training set should comprehensively cover those appearing in the target large system. To demonstrate DeepH's capability to generalize from small to large scales, we select a relaxed twisted bilayer MoTe$_2$ system containing 1,986 atoms as the prediction target. SOC is neglected here to keep the reference DFT calculation computationally tractable. To illustrate the significant impact of training-set design on model performance, we compare two different training-set generation strategies (Fig. \ref{fig5}(a)). Strategy I establishes a baseline dataset comprising 304 bilayer MoTe$_2$ structures divided into two categories. The first category consists of 256 supercells with 54 atoms, generated by systematically sampling in-plane interlayer translations on a uniform $16 \times 16$ grid. For these structures, the interlayer distance is uniformly sampled between 6.9 $\AA$ and 7.9 $\AA$ to capture the interlayer corrugation inherent in the relaxed twisted bilayer, and random atomic displacements within a range of $\pm 0.1$ $\AA$ are applied to enhance structural diversity. The second category targets the specific stacking motifs found in twisted systems. It includes 48 structures derived from relaxed twisted bilayers with unit cell sizes of 42, 114, and 222 atoms, which naturally cover mixed AA and AB stacking regions. Inspired by the MLIP validation protocol introduced in Ref.~\cite{georgaras2025accurate}, strategy II extends Strategy I by incorporating an additional set of 64 quasi-1D moir\'e structures, each containing 93 atoms. These structures are constructed by applying uniaxial strain to one layer relative to the other. This construction method allows the formation of rich moir\'e interference patterns within a supercell of moderate size, enabling the model to learn complex long-range modulations efficiently. Both strategies employed the same model architecture and hyperparameters: an embedding layer with 8.0 \r{A} Gaussian smearing, an intermediate representation of ``64×0e+48×1e+32×2e+32×3e+16×4e'', and a 4‑layer graph neural network with 2 attention heads per layer, resulting in 12,099,170 trainable parameters. All models were trained in FP32 precision on a single RTX 4090 GPU using a batch size of 1, the AdamW optimizer (initial learning rate = 0.001, \(\beta\) = (0.9, 0.999)), and a ReduceLROnPlateau scheduler with a patience of 60 epochs. For Strategy I, the 304 available structures were split into 228 for training, 38 for validation, and 38 for testing. Training converged after 1,111 epochs (11.1 hours), achieving final MAE of 0.091 meV on the training set, 0.105 meV on the validation set, and 0.085 meV on the test set. In contrast, Strategy II used 368 structures partitioned into 276 for training, 46 for validation, and 46 for testing. This model converged after 1,148 epochs (15.6 hours), with final MAEs of 0.084 meV (training), 0.115 meV (validation), and 0.120 meV (test).

For the 1,986-atom twisted bilayer MoTe$_2$, Strategy II significantly improves model accuracy over Strategy I: the MAE of the Hamiltonian matrix elements decreases  from 0.42 meV to 0.29 meV. This improvement is further reflected in band-structure predictions. As shown in Fig. \ref{fig5} (b,c), the band MAEs are 4.64 meV and 1.26 meV for Strategy I and Strategy II, respectively, with the accuracy of moir\'e flat bands showing particularly notable gains. The elimination of the deviations observed in Fig. \ref{fig5}(b) by the enhanced dataset in Fig. \ref{fig5}(c) attests to the high expressivity of the DeepH model. This contrast suggests that performance limitations are often attributable to data coverage rather than model capacity; therefore, systematically enriching the training set with representative local environments offers a reliable way to resolve potential accuracy bottlenecks.

\subsection{Amorphous silicon}

\begin{figure}
    \centering
    \includegraphics[width=\linewidth]{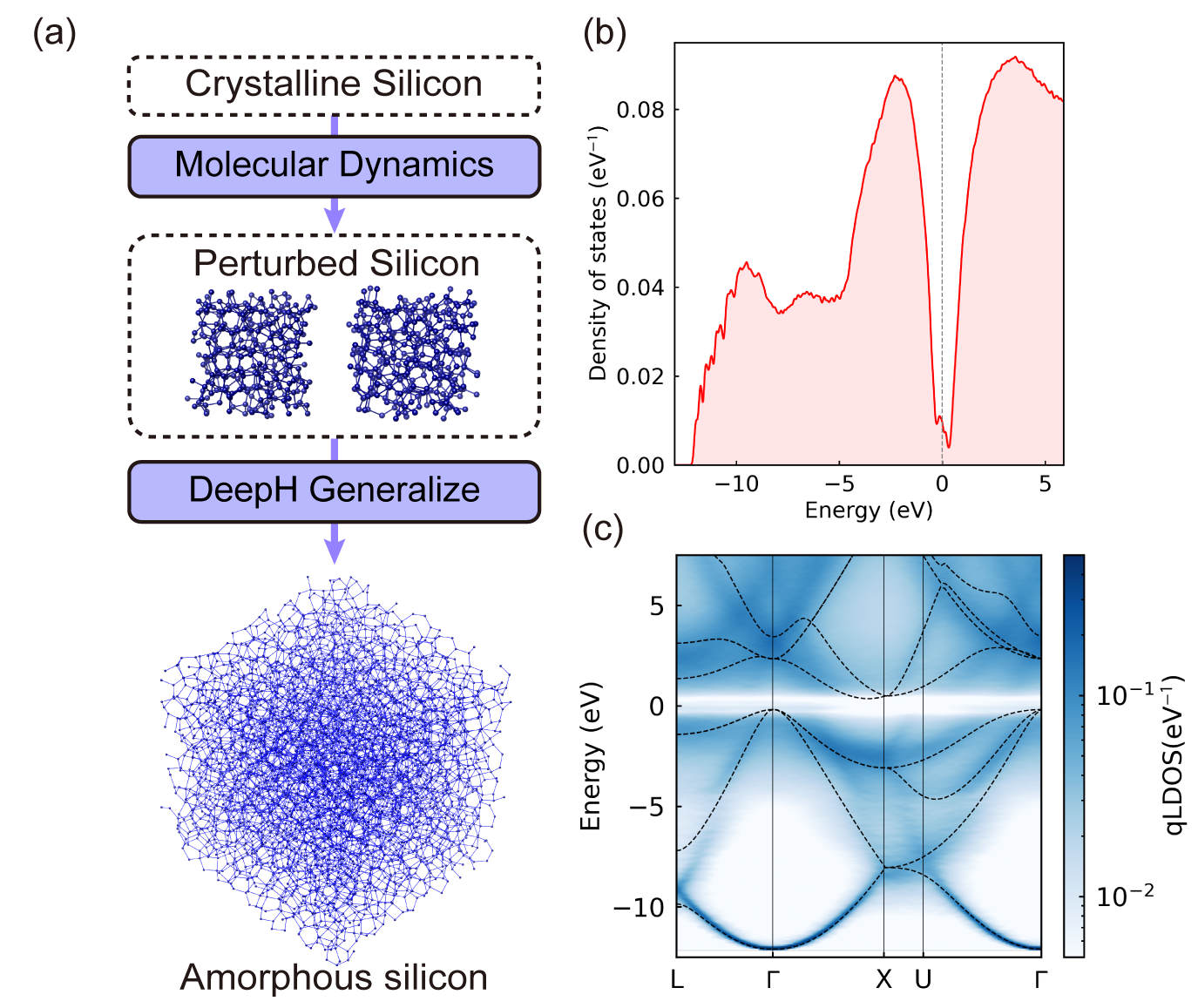}
    \caption{DeepH‑based electronic structure prediction for large‑scale amorphous silicon (a‑Si). (a) Workflow for generating the training data. 256 independent a‑Si configurations, each containing 216 atoms, were produced via melt‑quench molecular dynamics using the Stillinger‑Weber potential. The corresponding DFT‑calculated Hamiltonians serve as training labels. (b) Density of states (DOS) predicted by DeepH for an a‑Si system containing 4,096 atoms. (c) Effective band structure of the 4,096-atom a-Si system unfolded onto the Brillouin zone of FCC crystalline silicon. The plot is visualized as a heat map of the local density of states in reciprocal space (qLDOS), representing the momentum-resolved spectral weight. The black dashed lines correspond to the band structure of single-crystal silicon.}
    \label{fig6}
\end{figure}

Amorphous silicon (a-Si) serves as a prototypical model for the extensive family of non-crystalline semiconductors, finding critical applications in thin-film transistors and solar cells~\cite{stuckelberger2017progress,mirshojaeian2021review}. Unlike its crystalline counterpart, a-Si lacks long-range translational symmetry but retains a rigid short-range tetrahedral network. This structural duality presents a unique challenge for computational modeling. While DFT provides accurate descriptions of local bonding, the study of disorder-induced phenomena requires large supercells containing thousands of atoms to obtain statistically meaningful results. Such scales are often computationally intractable for standard DFT, hindering systematic investigations of structure-property relationships. However, since the electronic structure of a-Si is predominantly governed by local chemical environments and short-range covalent interactions, this material represents an ideal candidate system for applying the DeepH method. By accurately capturing this local physics from moderate-sized training samples, DeepH offers a scalable solution that maintains quantum-mechanical accuracy, effectively opening the door to efficient electronic-structure calculations for the broader class of amorphous and disordered materials.

To demonstrate this capability, we established a workflow to model the electronic properties of large-scale a-Si systems. To train a DeepH model, we first sample small a-Si structures via a melt-quench molecular dynamics procedure for training set construction. We used the Stillinger-Weber potential~\cite{vink2001fitting} implemented in the LAMMPS package~\cite{thompson2022lammps}. The simulation employed an NPT ensemble with a time step of 1 fs and followed a four-stage protocol: heating the system from 300 K to 2500 K over 30 ps, equilibrating at 2500 K for 1 ns, cooling down to 300 K over 2.2 ns, and finally equilibrating at 300 K for 150 ps. 300 independent a-Si configurations were generated from this process, each containing 216 atoms, to ensure a comprehensive sampling of the diverse local atomic environments found in the disordered network. For the training data, we performed self-consistent DFT calculations on these structures using Quantum ESPRESSO with the PBE functional and the optimized norm-conserving Vanderbilt pseudopotential (ONCVPSP)~\cite{hamann2013optimized}. The PW wavefunction cutoff was set to 60 Ry, and the k-point density was $3\times3\times3$. The resulting Hamiltonians were then projected from the PW basis onto SIESTA's Double-Zeta Polarized (DZP) basis set. As illustrated in Fig. \ref{fig6}(a), these labeled datasets were used to train the DeepH model. The network, based on the \texttt{Accurate} architecture, was trained on the dataset. The 300 structures were split into 240 for training, 30 for validation, and 30 for testing. The model was trained using FP32 precision over 11.3 hours on four RTX 4090 GPUs. Key hyperparameters include: an embedding Gaussian smearing of 7.0 \r{A}, an intermediate representation of ``48x0e+48x1e+40x2e+16x3e+8x4e'', a 4-layer GNN with 2 attention heads per layer, resulting in a total of 6,073,554 parameters. We used a batch size of 4 (four materials per batch), the AdamW optimizer with an initial learning rate of 0.001 and $\beta$ parameters =  (0.9, 0.999), and a ReduceLROnPlateau scheduler with a patience of 60 epochs and reduction factor of 0.5. The model was trained for 377 epochs, achieving a final training MAE of 0.58 meV, a validation MAE of 0.99 meV, and a test loss of 0.94 meV in Hamiltonian matrix elements.

Once trained, the model was deployed to predict the Hamiltonian of a significantly larger supercell containing 4096 atoms. This large-scale structure was adopted from Ref.~\cite{zongo2025amorphous}. To validate the physical accuracy of the prediction, we analyzed the spectral properties of this large-scale system. Figure \ref{fig6}(b) displays the predicted DOS, and Fig. \ref{fig6}(c) presents the effective band structure obtained via the band unfolding technique using SIESTA~\cite{mayo2020band,soler2002siesta}. The results successfully reproduce the characteristic features of the a-Si electronic structure, including disorder-induced broadening of energy bands, thereby confirming the model's robust transferability and its ability to bridge the scale gap in amorphous materials simulation. Our workflow integrates a series of computational tools: LAMMPS for structure generation, Quantum ESPRESSO for training-set production, DeepH-pack for model training, and SIESTA for post-processing. It demonstrates the flexibility of the DeepH framework in enabling seamless interoperability across software ecosystems. This effort highlights a step toward building a more unified and cohesive computational materials science community.

\subsection{Universal materials model}
Accurate determination of electronic structures is fundamental to understanding the physicochemical properties of materials at the microscopic level. While traditional first-principles methods offer high accuracy, their steep scaling with system size often imposes prohibitive computational costs, thereby limiting their applicability in high-throughput screening and large-scale simulations. The applications presented in previous sections exemplify the ``specialized model" approach within the DeepH framework. These models are tailored to specific material systems, enabling high-precision investigations of intricate physical properties driven by complex structural variations within a fixed compositional space. In parallel with these focused applications, a universal materials model of electronic structure represents a complementary paradigm that prioritizes broad transferability. It aims to establish a generalizable mapping from atomic configurations to electronic Hamiltonians across arbitrary elemental combinations. Consequently, a single trained model can not only provide rapid predictions for diverse materials but also serve as a robust pre-trained foundation that can be efficiently fine-tuned for downstream tasks. Assuming sufficient accuracy and chemical coverage, such a model has the potential to fundamentally transform the landscape of computational materials science. By circumventing the need for repetitive ab initio calculations, it facilitates high-throughput virtual screening across vast chemical spaces and serves as an efficient engine for inverse materials design and simulations of complex defect or interface systems. Ultimately, this approach provides a transformative computational tool to accelerate the discovery and design of novel materials.

\begin{figure*}
    \centering
    \includegraphics[width=0.7\linewidth]{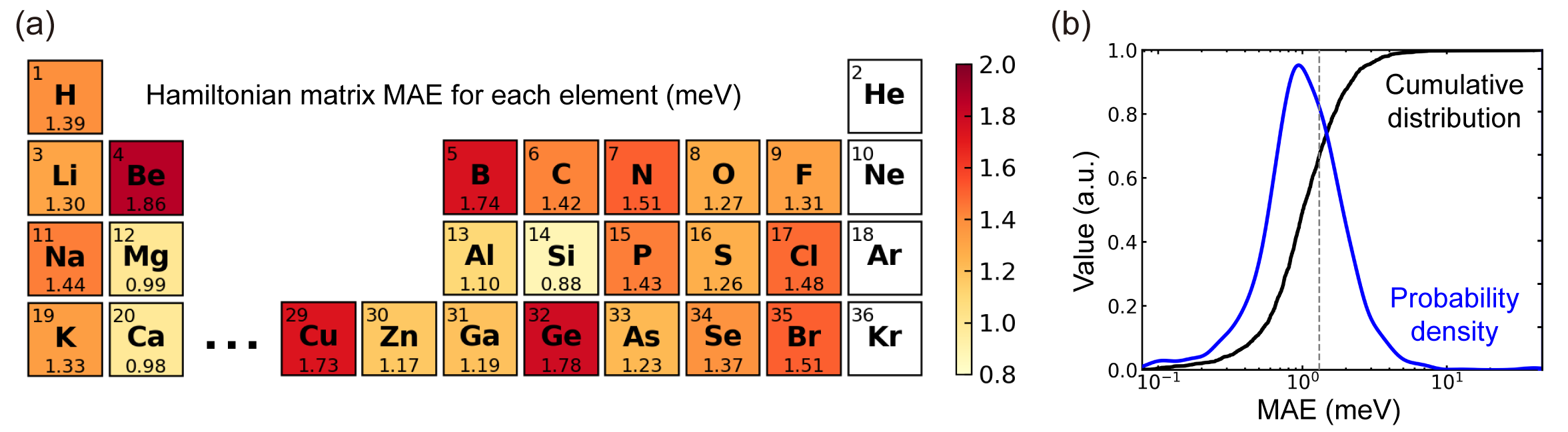}
    \caption{Prediction performance of the universal materials model of electronic structure. (a) Predicted mean absolute error (MAE) in Hamiltonian matrix for each element in the test set. (b) Distribution of prediction errors across the entire test set, showing a mean error of 1.31 meV.}
    \label{fig7}
\end{figure*}

The DeepH algorithm offers comprehensive support for training universal materials models of electronic structure. As a demonstration of this capability, we constructed an example universal materials model applicable to elements from the first four periods, focusing on non-magnetic systems without SOC. The training data were sourced from the Materials Project database, from which we selected  9,510 non-magnetic crystal structures composed of elements from the first four periods. The corresponding DFT Hamiltonian labels were calculated using the computational science workflow manager AiiDA~\cite{huber2020aiida,uhrin2021workflows} in combination with OpenMX~\cite{ozaki2003variationally}. Details on the dataset configuration can be found in the DeepH universal materials model publication~\cite{wang2024universal}. This dataset exhibits high diversity in element types, stoichiometry, crystal structure types, and bonding characteristics, ensuring the model can adequately learn the complex relationships between interatomic interactions and the electronic Hamiltonian across varied chemical environments~\cite{wang2024universal}.

The core architecture of the model is specifically designed to uniformly handle different element types and atomic systems of variable sizes. Its learning objective is to directly infer the DFT Hamiltonian from the input atomic coordinates and element types, from which various electronic properties can be derived. The network, based on the \texttt{Accurate} architecture, was trained on the dataset. The 9,510 structures were split into 5,692 for training, 1,927 for validation, and 1,891 for testing. The model was trained using FP32 precision over 99.6 hours on four RTX 4090 GPUs. Key hyperparameters include: an embedding Gaussian smearing of 9.0 \r{A}, an intermediate representation of ``96x0e+96x1e+80x2e+40x3e+20x4e+8x5e'', a 6-layer GNN with 2 attention heads per layer, resulting in a total of 47,761,367 parameters. We used a batch size of 1 (one material per batch), the AdamW optimizer with an initial learning rate of 0.0004 and $\beta$ parameters =  (0.9, 0.999), and a WarmupAndCosineDecay scheduler with 10,000 warmup steps. The model was trained for 151 epochs, achieving a final training MAE of 0.59 meV, a validation MAE of 1.27 meV, and a test loss of 1.31 meV in Hamiltonian matrix elements. The observed overfitting suggests there is room for further model improvement. By training on this large-scale, diverse dataset, the model does not merely memorize specific material instances. Instead, it internalizes the underlying physical rules connecting atomic numbers, geometric configurations, and quantum mechanical operators, thereby acquiring the capability to generalize and predict properties for previously unseen material structures.

To validate the effectiveness and demonstrate the strong potential of the constructed universal materials model, we applied it to perform predictions on a series of unseen material systems with diverse structures. This universal materials model successfully predicts the DFT Hamiltonians for materials spanning the first four periods of the periodic table across various morphologies. As shown in Fig. \ref{fig7}(a), the model achieves a prediction accuracy of better than 2 meV for all elements included in the test set. Figure \ref{fig7}(b) demonstrates that the mean prediction error across the test set is as low as 1.31 meV. The predicted results exhibit excellent agreement with first-principles benchmark calculations, which confirms that the model has successfully captured the universal governing laws of electronic structures across disparate materials. This work demonstrates that the deep-learning-based universal materials model of electronic structure has reached a level of practical utility. Its capability to generalize to a broad spectrum of material systems without structure-specific retraining lays a solid foundation for building the next-generation intelligent computational platform for materials science, marking a crucial step toward the ultimate goal of efficient, accurate, and comprehensive materials design and discovery.

\section{Conclusion}
This work presents DeepH-pack, a unified software toolkit that bridges first‑principles DFT with deep learning to enable accurate, efficient, and generalizable electronic structure calculations. By integrating fundamental physical principles (such as locality and equivariance) into neural‑network architectures, DeepH‑pack achieves ab‑initio accuracy while accelerating calculations by orders of magnitude. The package provides standardized interfaces to major DFT codes, a flexible data specification for Hamiltonian and related quantities, and end‑to‑end workflows for model training and property prediction. It thus establishes a robust computational foundation for large‑scale materials simulation and AI‑driven discovery.

The development of DeepH‑pack is built upon a series of methodological advances in deep‑learning electronic structure. Beginning with the DeepH framework, which learns Hamiltonian matrices from DFT data to predict material properties ~\cite{li2022deep}, the DeepH team systematically advanced the approach along two interconnected fronts: toward greater intelligence and toward greater generality. We introduced symmetry‑aware variants DeepH‑E3~\cite{gong2023general} and xDeepH~\cite{li2023deep} through equivariant neural networks, developed the Transformer‑based DeepH‑2 for enhanced transferability~\cite{wang2024deeph}, and constructed a universal DeepH materials model~\cite{wang2024universal}. The recent DeepH‑Zero method further demonstrated Hamiltonian optimization with zero training data by directly encoding physical constraints~\cite{li2024neural}. In parallel, the framework has been extended to support density functional perturbation theory calculations~\cite{li2024deep}, hybrid functional predictions~\cite{tang2023efficient}, plane‑wave program interfaces~\cite{gong2024generalizing}, density‑matrix learning~\cite{tang2024improving}, as well as being vastly exemplified on moir\'e-twisted materials~\cite{bao2024deep}, significantly broadening its applicability and physical completeness. Contributions from the broader DeepH community have further expanded the methodology's reach, supporting applications in non-adiabatic dynamics~\cite{liu2024breaking}, moir\'e heterostructures~\cite{yang2024evolution}, complex nanotube structures~\cite{huang2025atlas}, and defect systems~\cite{zhu2025predicting}, alongside new interfaces to additional DFT programs~\cite{ke2025combining}.

Looking ahead, DeepH is envisioned as an increasingly intelligent and universal platform for first‑principles electronic structure simulation. As the core implementation, DeepH-pack will be continuously refined to better serve the community, with focused improvements in usability, interoperability, and computational scalability. This ongoing development aims to empower researchers by facilitating a seamless transition into the AI-driven era of computational physics. Ultimately, by making high-accuracy electronic structure modeling simultaneously efficient, general, and accessible, DeepH-pack strives to fundamentally reshape the materials research ecosystem, accelerating the discovery and design of functional materials across scales.

\section*{Acknowledgements}
Yang Li, Boheng Zhao, Xiaoxun Gong, Yuxiang Wang, Zechen Tang, Zixu Wang, Zilong Yuan, Jialin Li, Minghui Sun, Zezhou Chen, Honggeng Tao, Baochun Wu, He Li, Wenhui Duan, and Yong Xu were supported by the Basic Science Center Project of NSFC (grant no. 52388201), the National Key Basic Research and Development Program of China (grants no. 2024YFA1409100 and no. 2023YFA1406400), the National Natural Science Foundation of China (grants no. 12334003, no. 12421004, and no. 12361141826), the Fundamental and Interdisciplinary Disciplines Breakthrough Plan of the Ministry of Education of China (grant no. JYB2025XDXM408), ShanghaiTech AI4S Initiative SHTAI4S202504, the National Science Fund for Distinguished Young Scholars (grant no. 12025405), the China Postdoctoral Science Foundation (grant no. 2025M773367), and the NSFC Basic Research Scheme for PhD Students (grant no. 124B2072). Yanzhen Wang and Felipe H. da Jornada acknowledge support by the National Science Foundation (NSF) CAREER award through Grant No. DMR-2238328.


\end{document}